\documentclass[10pt,journal,compsoc]{IEEEtran}

\usepackage[T1]{fontenc}
\usepackage[utf8]{inputenc}
\usepackage{mathtools}
\usepackage[colorlinks,urlcolor=blue,linkcolor=blue,citecolor=blue]{hyperref}
\usepackage[style=ieee]{biblatex}
\usepackage{listings}
	\lstdefinelanguage{diff}{
    basicstyle=\ttfamily\small,
    morecomment=[f][\color{diffstart}]{@@},
    morecomment=[f][\color{diffincl}]{+},
    morecomment=[f][\color{diffrem}]{-},
  }
\usepackage[svgnames]{xcolor}
  \definecolor{diffstart}{named}{Grey}
  \definecolor{diffincl}{named}{Green}
  \definecolor{diffrem}{named}{OrangeRed}
\usepackage{xspace}
\usepackage[inline]{enumitem}
\usepackage{caption} 
\usepackage{amssymb}
\usepackage{numprint}
\usepackage{tabu}
\usepackage{multirow}
\usepackage{booktabs} % For formal tables
\usepackage{siunitx}
\usepackage{amsmath}
\usepackage{mdframed}
\usepackage{makecell}

\newcommand{\ie}{\textit{i.e.,}\xspace}
\newcommand{\eg}{\textit{e.g.,}\xspace}
\newcommand{\etc}{\textit{etc.}\xspace}
\newcommand{\etal}{\textit{et al.}\xspace}

\newcommand\minor[1]{\textcolor{black}{#1}}

\bibliography{references.bib}

\begin{document}
\title{Neural Transfer Learning for Repairing\\ Security Vulnerabilities in C Code}

\author{Zimin~Chen,
        ~Steve~Kommrusch,
        and~Martin~Monperrus% <-this % stops a space
\IEEEcompsocitemizethanks{\IEEEcompsocthanksitem Zimin Chen and Martin Monperrus are with KTH Royal Institute of Technology, Sweden.\protect\\
E-mail: \{zimin, monp\}@kth.se
\IEEEcompsocthanksitem Steve Kommrusch is with Colorado State University, USA.\protect\\
E-mail: steveko@cs.colostate.edu
\IEEEcompsocthanksitem Zimin Chen and Steve Kommrusch have equally contributed to the paper as first authors.
}% <-this % stops an unwanted space
\thanks{Manuscript submitted 2021-04-16.}}

\IEEEtitleabstractindextext{%
\begin{abstract}
In this paper, we address the problem of automatic repair of software vulnerabilities with deep learning. The major problem with data-driven vulnerability repair is that the few existing datasets of known confirmed vulnerabilities consist of only a few thousand examples. However, training a deep learning model often requires hundreds of thousands of examples. In this work, we leverage the intuition that the bug fixing task and the vulnerability fixing task are related and that the knowledge learned from bug fixes can be transferred to fixing vulnerabilities. In the machine learning community, this technique is called transfer learning. In this paper, we propose an approach for repairing security vulnerabilities named VRepair which is based on transfer learning. VRepair is first trained on a large bug fix corpus and is then tuned on a vulnerability fix dataset, which is an order of magnitude smaller. In our experiments, we show that a model trained only on a bug fix corpus can already fix some vulnerabilities. Then, we demonstrate that transfer learning improves the ability to repair vulnerable C functions. We also show that the transfer learning model performs better than a model trained with a denoising task and fine-tuned on the vulnerability fixing task. To sum up, this paper shows that transfer learning works well for repairing security vulnerabilities in C compared to learning on a small dataset.
\end{abstract}

}

\maketitle

\IEEEdisplaynontitleabstractindextext

% \fi
\IEEEpeerreviewmaketitle

\IEEEraisesectionheading{\section{Introduction}\label{sec:introduction}}

\IEEEPARstart{O}{n} the code hosting platform GitHub, the number of newly created code repositories has increased by 35\% in 2020 compared to 2019, reaching 60 million new repositories during 2020 \cite{Octoverse}. This is a concern to security since the number of software security vulnerabilities is correlated with the size of the software \cite{radjenovic2013software}. Perhaps worryingly, the number of software vulnerabilities is indeed constantly growing \cite{homaei2017seven}. Manually fixing all these vulnerabilities is a time-consuming task; the GitHub 2020 security report finds that it takes 4.4 weeks to release a fix after a vulnerability is identified \cite{OctoverseSecurity}. Therefore, researchers have proposed approaches to automatically fix these vulnerabilities \cite{harer2018learning, chi2020seqtrans}.

In the realm of automatic vulnerability fixing \cite{ji2018coming}, there are only a few works on using neural networks and deep learning techniques. One of the reasons is that deep learning models depend on acquiring a massive amount of training data \cite{sun2017revisiting}, while the amount of confirmed and curated vulnerabilities remains small. Consider the recent related work Vurle \cite{ma2017vurle}, where the model is trained and evaluated on a dataset of 279 manually identified vulnerabilities. On the other hand, training neural models for a translation task (English to French) has been done using over 41 million sentence pairs \cite{bojar-EtAl:2014:W14-33}. Another example is the popular summarization dataset CNN-DM  \cite{hermann2015teaching} that contains 300 thousand text pairs. Li \etal showed that the knowledge learned from a small dataset is unreliable and imprecise \cite{li2007using}.
Schmidt \etal found that the error of a model predicting the thermodynamic stability of solids decreases with the size of training data \cite{schmidt2017predicting}. We argue that learning from a small dataset of vulnerabilities suffers from the same problems (and will provide empirical evidence later). 

In this paper, we address the problem that vulnerability fix datasets are too small to be meaningfully used in a deep-learning model. Our key intuition to mitigate the problem is to use transfer learning. Transfer learning is a technique to transfer knowledge learned from one domain to solve problems in related domains, and it is often used to mitigate the problem of small datasets \cite{adams2017cross}. We leverage the similarity of two related software development tasks: bug fixing and vulnerability fixing. In this context, transfer learning means acquiring generic knowledge from a large bug fixing dataset and then transferring the learned knowledge from the bug fixing task to the vulnerability fixing task by tuning it on a smaller vulnerability fixing dataset. We realize this vision in a novel system for automatically repairing C vulnerabilities called VRepair.

To train VRepair, we create a sizeable bug fixing dataset, large enough to be amenable to deep learning. 
We create this dataset by collecting and analyzing all 892 million GitHub events that happened between 2017-01-01 and 2018-12-31 and using a heuristic technique to extract all bug fix commits. In this way, we obtain a novel dataset consisting of over 21 million bug fixing commits in C code.
We use this data to first train VRepair on the task of bug fixing. Next, we use two datasets of vulnerability fixes from previous research, called Big-Vul \cite{fan2020ac} and CVEfixes \cite{bhandari2021cvefixes}. We tune VRepair on the vulnerability fixing task based on both datasets. Our experimental results show that the bug fixing task can be used to train a model meant for vulnerability fixing; the model trained only on the collected bug fix corpus achieves \minor{18.24\% accuracy on Big-Vul and 15.98\% on CVEfixes}, which validates our initial intuition that these tasks are profoundly similar. Next, we show that by using transfer learning, \ie by first training on the bug fix corpus and then by tuning the model on the small vulnerability fix dataset, VRepair increases its accuracy to \minor{21.86\% on Big-Vul and 22.73\% on CVEfixes}, demonstrating the power of transfer learning. Additionally,  we compare transfer learning with a denoising pre-training followed by fine-tuning on the vulnerability fixing task and show that VRepair's process is better than pre-training with a generic denoising task. We also show that transfer learning improves the stability of the final model.

In summary, our contributions are:

\begin{itemize}
% Old first item: \item We design a novel code representation for the program repair task with neural networks. Our output code representation is a token difference instead of the whole fixed source code used in recent research \cite{tufano2019empirical, gupta2017deepfix}. This code representation is used to train a deep learning Transformer in a system called VRepair.

% core conceptual contribution
\item We introduce VRepair, a Transformer Neural Network Model which targets the problem of vulnerability repair. The core novelty of VRepair is to employ transfer learning as follows: it is first trained on a bug fix corpus and then tuned on a vulnerability fix dataset.

% experiments
\item We empirically demonstrate that on the vulnerability fixing task, the transfer learning VRepair model performs better than the alternatives:
1) VRepair is better than a model trained only on the small vulnerability fix dataset;
2) VRepair is better than a model trained on a large generic bug fix corpus;
3) VRepair is better than a model pre-trained with a denoising task.
In addition, we present evidence that the performance of the model trained with transfer learning is stable.

% open science
\item We share all our code and data for facilitating replication and fostering future research on this topic.
\end{itemize}

% \item We design a novel code representation for the program repair task with neural networks. Our output code representation is a token difference instead of the entire fixed source code used in recent research \cite{tufano2019empirical, gupta2017deepfix}.

\section{Background}

\subsection{Software Vulnerabilities}

A software vulnerability is a weakness in code that can be exploited by an attacker to perform unauthorized actions. For example, one common kind of vulnerability is a buffer overflow, which allows an attacker to overwrite a buffer's boundary and inject malicious code. Another example is an SQL injection, where malicious SQL statements are inserted into executable queries. The exploitation of vulnerabilities contributes to the hundreds of billions of dollars that cybercrime costs the world economy each year \cite{losses2014estimating}.

Each month, thousands of such vulnerabilities are reported to the Common Vulnerabilities and Exposures (CVE) database. Each one of them is assigned a unique identifier.
Each vulnerability with a CVE ID is also assigned to a Common Weakness Enumeration (CWE) category representing the generic type of problem. 

\textbf{Definition} a \emph{CVE ID} identifies a vulnerability within the Common Vulnerabilities and Exposures database. It is a unique alphanumeric assigned to a specific vulnerability.

For instance, the entry identified as CVE-2019-9208 is a vulnerability in Wireshark due to a null pointer exception. 2019 is the year in which the CVE ID was assigned or the vulnerability was made public; 9208 uniquely identifies the vulnerability within the year.

\textbf{Definition} a \emph{CWE ID} is a Common Weakness Enumeration and identifies the category that a CVE ID is a part of. CWE categories represent general software weaknesses.

For example CVE-2019-9208 is categorized into CWE-476, which is the 'NULL Pointer Dereference' category. In 2020, CWE-79 (Cross-site Scripting), CWE-787 (Out-of-bounds Write) and CWE-20 (Improper Input Validation) are the 3 most common vulnerability categories\footnote{\url{https://cwe.mitre.org/top25/archive/2020/2020_cwe_top25.html}}. 

Each vulnerability represents a threat until a patch is written by the developers.

\subsection{Transformers}

Sequence-to-sequence (seq2seq) learning is a modern machine learning framework that is used to learn mappings between two sequences, typically of words \cite{sutskever2014sequence}. It is widely used in automated translation \cite{wu2016google}, text summarization \cite{nallapati2016abstractive} and other tasks related to natural language. A seq2seq model consists of two parts, an encoder and a decoder. The encoder maps the input sequence $X = (x_{0},x_{1},...,x_{n})$ to an intermediate continuous representation $H = (h_{0},h_{1},...,h_{n})$. Then, given $H$, the decoder generates the output sequence $Y = (y_{0}, y_{1},...,y_{m})$. Note that the size of the input and output sequences, $n$ and $m$, can be different. A seq2seq model is optimized on a training dataset to maximize the conditional probability of $p(Y \mid X)$, which is equivalent to:

\begin{align*}
\begin{split}
p(Y \mid X) &= p(y_{0},y_{1},...,y_{m} \mid x_{0},x_{1},...,x_{n})\\
            &= \prod_{i=0}^{m} p(y_{i} \mid H, y_{0},y_{1},...,y_{i-1})
\end{split}
\end{align*}

Prior work has shown that source code has token and phrase patterns that are as natural as human language \cite{hindle2012naturalness}, and thus techniques used in natural language processing can work on source code as well, including seq2seq learning \cite{chen2019sequencer}. 
\begin{figure*}[!t]
    \centering
    \includegraphics[width=0.95\textwidth]{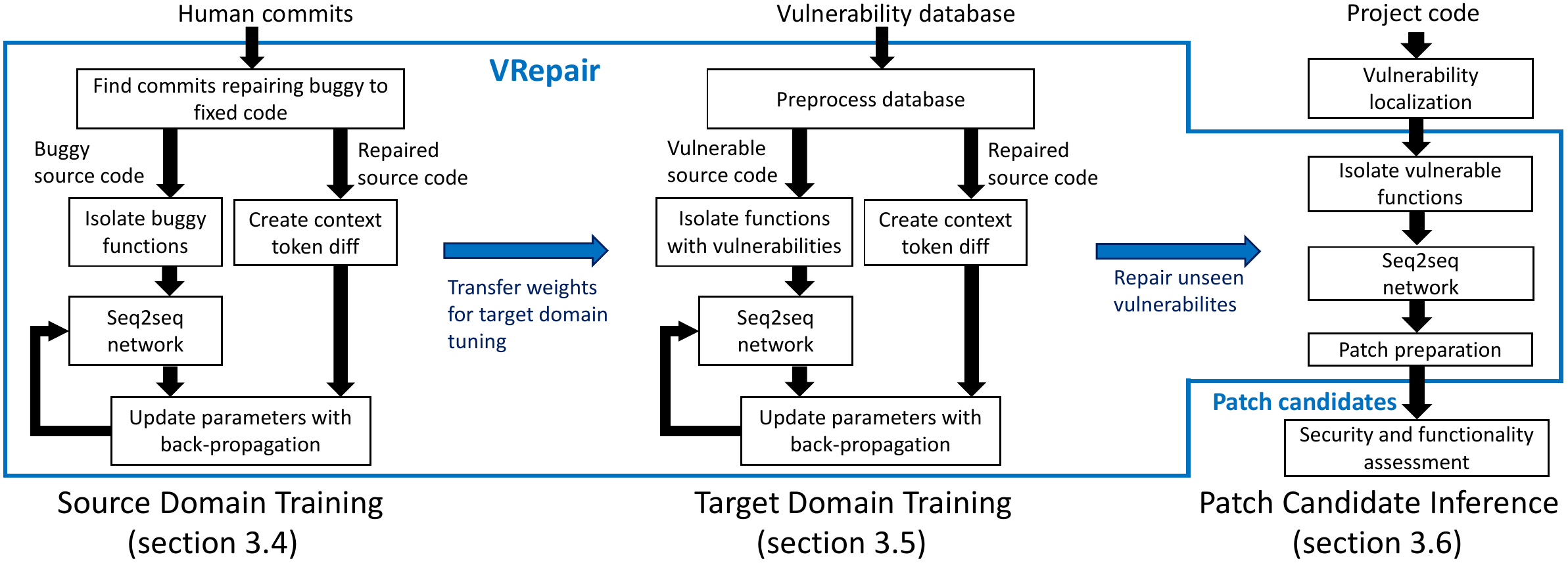}
    \caption{The VRepair Workflow: Two phases of training for transfer learning to repair vulnerabilities.}
    \label{fig:workflow}
\end{figure*}

In our work, we use a variant of a seq2seq model called the ``Transformer'' \cite{vaswani2017attention}, which is considered the state-of-the-art architecture for seq2seq learning. 
The main improvement of Transformers is the usage of self-attention. The number of operations required to propagate information between two tokens in seq2seq learning based on recurrent neural networks grows linearly with the distance, while in the Transformer architecture it is a constant time operation with the help of self-attention. Self-attention updates all hidden states simultaneously by allowing all tokens to attend to all previous hidden states. Since there are no dependencies between tokens, this operation can be done in parallel without recurrence. This also helps to mitigate the issue of long term dependencies which can be a problem for recurrent neural networks. Self attention can be described as the process of mapping a query and key-value pairs to an output. Query, key, and value are concepts borrowed from information retrieval systems, where the search engine maps a query against keys and returns values. In a Transformer model, Query (Q), key (K), and value (V) are vectors of the same dimension $d_{k}$ computed for each input. The attention function is computed as:

$$
\text{Attention}(Q, K, V) = \text{softmax}(\frac{QK^{T}}{\sqrt{d_{k}}})V
$$

The Transformer model is trained with multiple attention functions (called multi-head attention) which allows each attention function to attend to different learned representations of the input. Since the Transformer model computes all hidden states in parallel, it has no information about the relative or absolute position of the input. Therefore the Transformer adds positional embedding to the input embeddings, which is a vector representing the position.

The encoder of a Transformer has several layers, each layer having two sub-layers. The first sub-layer is a multi-head self-attention layer, and the second sub-layer is a feed forward neural network. The outputs from both sub-layers are normalized using layer normalization together with residual connections around each sub-layer.

The decoder also has several layers, each layer has three sub-layers. Two of the decoder sub-layers are similar to the two sub-layers in the encoder layer, but there is one more sub-layer which is a multi-head self-attention over the output of the encoder. The Transformer architecture we utilize is shown in \autoref{fig:transformer} and discussed in further detail in \autoref{sec:architecture}.

\subsection{Transfer Learning}
\label{sec:transfer-learning}
Traditional machine learning approaches learn from examples in one domain and solve the problem in the same domain (\eg learning to recognize cats from a database of cat images).
However, as humans, we do better: we are able to apply existing knowledge from other relevant domains to solve new tasks. The approach works well when two tasks share some commonalities, and we are able to solve the new problem faster by starting at a point using previously learned insights. Transfer learning is the concept of transferring knowledge learned in one source task to improve learning in a related target task \cite{torrey2010transfer}. The former is called the \textit{source domain} and the latter the \textit{target domain}. Transfer learning is commonly used to mitigate the problem of insufficient training data \cite{adams2017cross}. If the training data collection is complex and hard for a target task, we can seek a similar source task where massive amounts of training data are available. Then, we can train a machine learning model on the source task with sufficient training data, and tune the trained model on the target task with the limited training data that we have.

Tan \etal divide transfer learning approaches into four categories: 1) Instance-based 2) Mapping-based 3) Adversarial-based 4) Network-based   \cite{tan2018survey}. Instance-based transfer learning is about reusing similar examples in the source domain with specific weights as training data in the target domain. Mapping-based transfer learning refers to creating a new data space by transforming inputs from both the source and target domains into a new representation. Adversarial-based transfer learning refers to using techniques similar to generative adversarial networks to find a representation that is suitable on both the source and target domain. Network-based transfer learning is where we reuse the network structure and parameters trained on the source domain and transfer the knowledge to the target domain; this is what we do in this paper.

\minor{In this paper, we choose to use network-based transfer learning as the foundation for VRepair because bug repair data is readily available for pretraining a network which can then be incrementally trained on the target task. The target problem of vulnerability repair already maps into a sequence-to-sequence model well and the source training data for buggy program repair can use the same representation - the representation of attention layers in a transformer model. For instance-based learning, a subset of GitHub bug fixes could potentially be used by selecting relevant samples but this adds complexity to the use model. The mapping-based approach for transfer learning is not an obvious fit for our sequence-to-sequence problem. Finally, using an adversarial network does not fit well with our problem space primarily because our system does not include an automated mechanism to evaluate (test) alternative outputs.}

\section{VRepair: Repairing Software Vulnerabilities with Transfer Learning}

In this section, we present a novel neural network approach to automatically repair software vulnerabilities.

\subsection{Overview}

VRepair is a neural model to fix software vulnerabilities, based on the state-of-the-art Transformer architecture. The prototype implementation targets C code and is able to repair intraprocedural vulnerabilities in C functions. VRepair uses transfer learning (see \autoref{sec:transfer-learning}) and thus is composed of three phases: source domain training, target domain training, and inference, as shown in \autoref{fig:workflow}.

\textbf{Source domain training} is our first training phase. We train VRepair using a bug fix corpus because it is relatively easy to collect very large numbers of bug fixes by collecting commits (\eg on GitHub). While this corpus is not specific to vulnerabilities, per the vision of transfer learning, VRepair will be able to build some knowledge that would turn valuable for fixing vulnerabilities.
From a technical perspective, training on this corpus sets the neural network weights of VRepair to some reasonable values with respect to manipulating code and generating source code patches in the considered programming language.

\textbf{Target domain training} is the second phase after the source domain training. In this second phase, we used a high quality dataset of vulnerability fixes. While this dataset only contains vulnerability fixes, its main limitation is its size because vulnerability fixes are notoriously scarce. Based on this dataset, we further tune the weights in the neural network of VRepair. In this phase, VRepair transfers the knowledge learned from fixing bugs to fixing vulnerabilities. As we will demonstrate later in \autoref{subsec:result_rq2}, VRepair performs better with transfer learning than with just training on the small vulnerability fixes dataset.

\begin{figure*}[!t]
    \centering
    \includegraphics[width=1.0\textwidth]{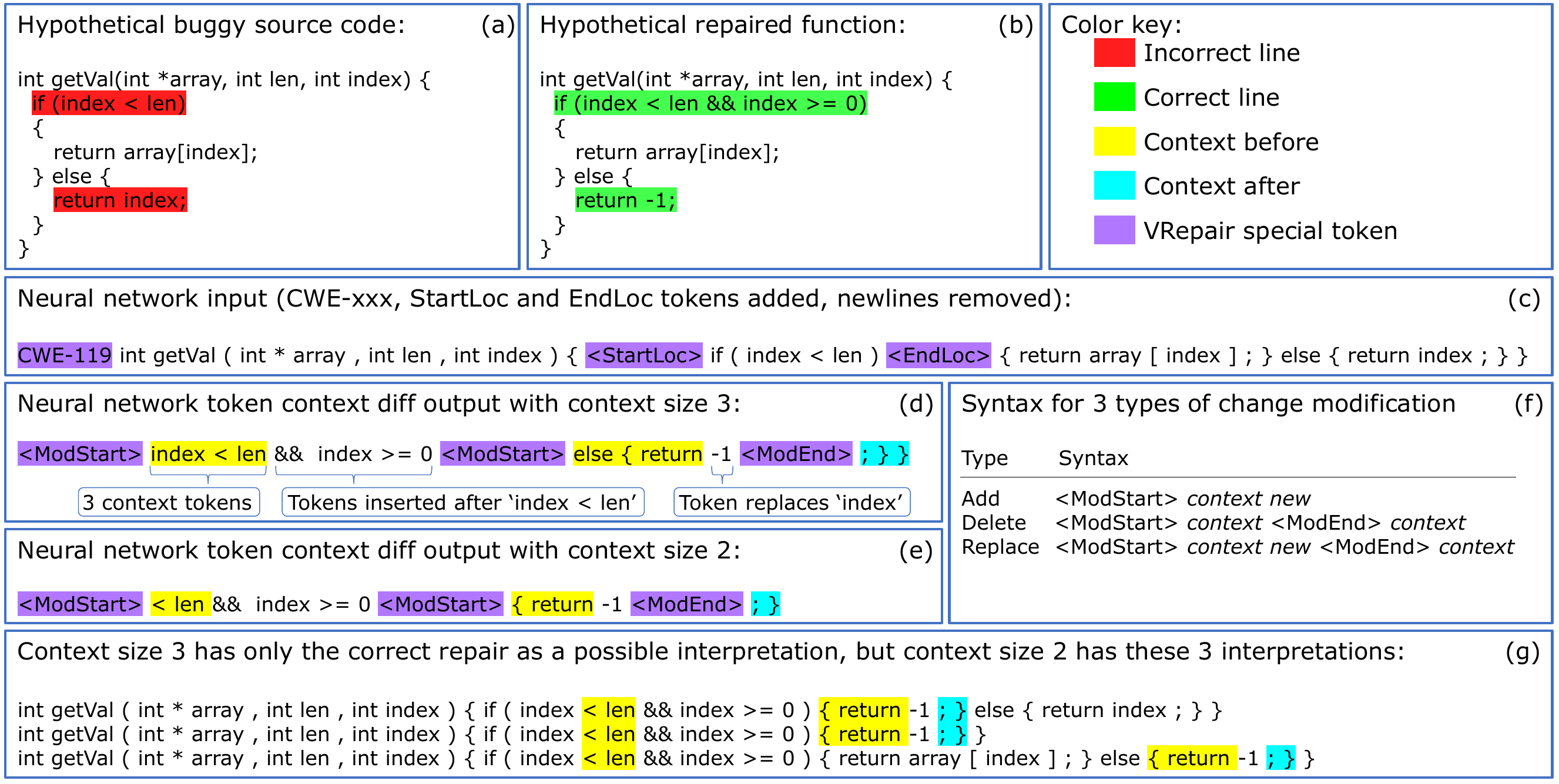}
    \caption{The VRepair code representation, based on a token diff script. Outputs are shown for a network trained to produce context size 3 and another network trained to produce context size 2.}
    \label{fig:context}
\end{figure*}

\textbf{Inference} is the final phase, where VRepair predicts vulnerability fixes on previously unseen functions known to be vulnerable according to a given vulnerability localization technique. VRepair is agnostic with respect to vulnerability localization, it could be for example a human security expert or a static analyzer dedicated to some vulnerability types.  
For each potentially vulnerable location, VRepair may output multiple predictions which may be transformable into multiple source code patches. Those tentative patches are meant to be verified in order to check whether the vulnerability has disappeared. This means that, like localization, verification is independent of VRepair. It would typically be another human security expert, or the same static analyzer used before.

\subsection{Code Representation}
\label{subsec:data_representation}

Early works on neural repair \cite{tufano2019empirical} simply output the full function for the proposed patched code, but the performance of such a system was seen to diminish as the function length increased \cite{tufano2019empirical}. Recently, other output representations have been proposed, including outputting a single line \cite{chen2019sequencer}, and outputting transformations as repair instructions \cite{mesbah2019deepdelta,tarlow2020learning,Zhu2021Repair}. Inspired by this recent related work, we develop a token-based change description for repairs. It allows for sequences shorter than a full function to be generated, it is easily applied to code, and it can represent any token change necessary for code repair. To sum up, in VRepair, the core idea is to represent the patch as an edit script at the token level.

The full overview of the input and output formats for VRepair are shown in \autoref{fig:context}. As shown in box (c), VRepair adjusts the input by removing all newline characters and identifying the first suspicious code line with the special tokens \texttt{<StartLoc>} and \texttt{<EndLoc>}. Those localization tokens come from an external entity localizing the vulnerable line, such as any vulnerability detection system from the corresponding prolific literature \cite{lin2020software,li2018vuldeepecker,russell2018automated} or a human security expert. For instance, a static analyzer such as Infer \cite{Infer} outputs a suspicious line that causes a vulnerability. The input function is also labeled with the vulnerability type suspected by adding a vulnerability type token at the start of the Transformer model input as detailed in \autoref{subsec:source_domain_training} (prefixed by CWE by convention).

Regarding the neural network output, it is known that a key challenge for using neural nets on code is that many functions are made of long token sequences \cite{tufano2019empirical}, and the performance of seq2seq models decreases with the size of the sequence \cite{cho2014properties}. In VRepair, the core idea is that the network only outputs the changed source code tokens, and not the whole function, \ie not the whole token sequence. 
By doing this:
1) the representation allows for representing multiple changes to a function
2) the representation decreases the size of the output sequence to be generated.
As seen in box (d) on \autoref{fig:context}, our system uses a special token to identify a change, \texttt{<ModStart>}, followed by $n_{context}$ tokens which locate where the change should start by specifying the tokens before a change. \texttt{<ModEnd>} is used when tokens from the input program are being replaced or deleted, and it is followed by $n_{context}$ tokens to specify the completion of a change.

\textbf{Definition} a \emph{token context diff} is a full change description for a function, identifying multiple change locations on multiple lines, using two special tokens \texttt{<ModStart>} and  \texttt{<ModEnd>} to identify the start and end contexts for change locations. 

\textbf{Definition} The \emph{context size} is the number of tokens from the source code used to identify the location where a change should be made. The context after the \texttt{<ModStart>} special token indicates where a change should begin, and the context after a  \texttt{<ModEnd>} special token indicates where a change that removes or replaces tokens should end.

%% example
For example, \autoref{fig:context} boxes (a) and (b) show a buggy function in which 2 lines are changed to repair the vulnerability. 
For this repair, the change encoded with a \emph{context size} of 3 results in a 17-token sequence show in box (d). Box (e) shows that the same change can be represented with a \emph{context size} of 2, this would result in a 14 token sequence.

%% how we represent Add Delete Replace
There are 3 types of changes used to patch code: new tokens may be added, tokens may be deleted, or tokens may be replaced. 
Our novel code representation supports them all, as shown in box (f) of \autoref{fig:context}.
Technically, only add and delete would be sufficient, but the replacement change simplifies the specification and allows a \emph{token context diff} to be processed sequentially without backtracking.

% discussion about context size
A shorter context size minimizes the output length to specify a code change, hence potentially facilitates the learning task. 
Yet, the issue that arises is that shorter \emph{context sizes} risk having multiple interpretations. For example, in box (d) of \autoref{fig:context}, the context size 3 specification uniquely identifies the start and end locations for the token modifications in the example function. The 3 token start contexts '\texttt{index < len}' and '\texttt{else \{ return}' and the end context '\texttt{; \} \}}' each only occur once in the buggy source code, so there is no ambiguity about what the resulting repaired function should be. But as shown in box (g), the context size 2 specification has multiple interpretations for the token replacement change. This is because '\texttt{\{ return}' and '\texttt{; \}}' both occur 2 times in the program resulting in 3 possible ways the output specification can be interpreted. For example, the first time the tokens '\texttt{\{ return}' occur in the buggy source code is directly before '\texttt{arrax[index]}', which is not the correct location to begin the modification to produce the correct repaired function. In this example, we see that a 2-token context is ambiguous whereas 3 tokens are sufficient to uniquely identify a single possible patch. Our pilot experiments showed that a context size of 3 successfully represents most patches without ambiguity, and hence we use this context size to further develop VRepair.

% multi-line explanation
In addition to confirming a context size of 3, our initial experiments have shown that the \emph{token context diff} approach supports multi-line fixes, which is an improvement on prior work only attempting to repair single lines \cite{chen2019sequencer}. We note that 49\% of our source domain dataset which we will use in the experiment (see \autoref{subsec:dataset}) contains multi-line fixes, demonstrating the importance of the \emph{token context diff}. We also present in \autoref{subsec:result_rq3} a case study of such a multi-line patch.

In summary, VRepair introduces a novel context-based representation for code patches, usable by a neural network to predict the location where specific tokens need to be added, deleted, or modified. Compared to other works which limit changes to single lines \cite{chen2019sequencer} or try to output the entire code of repaired functions \cite{gupta2017deepfix, tufano2019empirical}, our approach allows for complex multi-line changes to be specified with a small number of output tokens.

\subsection{Tokenization}

In this work we use a programming language tokenizer with no subtokenization.
We use Clang as the tokenizer because it is the most powerful tokenizer we are aware of, able to tokenize un-preprocessed C source code. We use the copy mechanism to deal with the out-of-vocabulary problem per \cite{chen2019sequencer}.
We do not use sub-tokenization such as BPE \cite{sennrich2015neural} because it increases the input and output length too much (per the literature \cite{cho2014properties}, confirmed in our pilot experiments).
Variable renaming is a technique that renames function names, integers, string literals, \etc to a pool of pre-defined tokens. For example function names \textit{GetResult} and \textit{UpdateCounter} can be replaced with FUNC\_1 and FUNC\_2. 
We do not use variable renaming because it hides valuable information about the variable that can be learned by word embeddings. For example, \textit{GetResult} should be a function that returns a result.

\subsection{Source Domain Training}
\label{subsec:source_domain_training}

In the source domain training phase, we use a corpus of generic bug fixes to train the model on the task of bug fixing. In this phase, the network learns about the core syntax and the essential primitives for patching code.
This works as follows.
From the bug fix corpus, we extract all functions that are changed. Each function is used as a training sample: the function before the change is seen as buggy, and the function after the change is the fixed version. 

We follow the procedure described in \autoref{subsec:data_representation} to process the buggy and fixed functions and extract the VRepair representation to be given to the network. All token sequences are preceded with a special token 'CWE-xxx', indicating what type of CWE category this vulnerability belongs to. We add this special token because we believe that vulnerabilities with the same CWE category are fixed in a similar way. For the bug fix corpus where we don't have this information, we use the 'CWE-000' token meaning ``generic fix''. This special token is mandatory for target domain training and inference as well, as we shall see later.

% Sentence at original submission: Per the usual practice \cite{prechelt1998early}, we apply early stopping during the source domain training. 
In machine learning, overfitting can occur when a model has begun to learn the training dataset too well and does not generalize well to unseen data samples. To combat this issue we use the common practice of early stopping \cite{prechelt1998early} during source domain training. For early stopping, a subset of our generic bug fix dataset is withheld for model validation during training. If the validation accuracy does not improve after two evaluations, we stop the source domain training phase and use the model with the highest validation accuracy for target domain training.

\subsection{Target Domain Training} \label{subsec:target_domain_training}

Next, we use the source domain validation dataset to select the best model produced by source domain training, and tune it on a vulnerability fixes dataset. The intuition is that the knowledge from bug fixing can be transferred to the vulnerability fix domain. We follow the same procedure mentioned in the \autoref{subsec:source_domain_training} to extract the buggy and fixed functions in the VRepair representation from all vulnerability fixes in the vulnerability dataset. We precede all inputs with a special token 'CWE-xxx' identifying the CWE category to which the vulnerability belongs. To ensure sufficient training data for each 'CWE-xxx' special token, we compute the most common CWE IDs and only keep the CWE IDs with sufficient examples in the vocabulary. The top CWE IDs that we keep cover 80\% of all vulnerabilities and the CWE IDs that are not kept in the vocabulary are replaced with 'CWE-000' instead. Early stopping is also used here, and the model with the highest validation accuracy is used for inference, \ie it is used to fix unforeseen vulnerabilities.

\subsection{Inference for Patch Synthesis}
\label{sec:inference}

After the source and target domain training, VRepair is ready to be used as a vulnerability fixing tool. The input provided to VRepair at inference time is the token sequence of a potentially vulnerable function with the first vulnerable line identified with special tokens \texttt{<StartLoc>} and \texttt{<EndLoc>}, as described in \autoref{subsec:data_representation}. The input should also be preceded by a special token 'CWE-xxx', representing the type of vulnerability. If the vulnerability was found by a static analyzer, we use a one-to-one mapping from each static analyzer warning to a CWE ID. 
% Original submission: The sequence-to-sequence model is then used with a special kind of beam search to generate multiple predictions, called the VRepair beam, defined next. 
VRepair then uses the Transformer model to create multiple \emph{token context diff} proposals for a given input. For each prediction of the neural network, VRepair finds the context to apply the patch and applies the predicted patch to create a patched function.

As we discuss in Section~\ref{sec:architecture}, the Transformer model learns to produce outputs that are likely to be correct based on the training data it has processed. Beam search is a well-known method \cite{tufano2019empirical} in which outputs from the model can be ordered and the $n$ most likely outputs can be considered by the system. Beam search is an important part of inference in VRepair. We introduce in VRepair a novel kind of beam, which is specific to the code representation introduced in \autoref{subsec:data_representation}, defined as follows. 
% Old wording from original submission: It is an extension of the well-known neural beam \cite{tufano2019empirical}.

\textbf{Definition} The \emph{Neural beam} is the set of predictions created by the neural network only, starting with the most likely one and continuing until the maximum number of proposals configured has been output \cite{tufano2019empirical}. Neural beam search works by keeping the $n$ best sequences up to the current decoder state. The successors of these states are computed and ranked based on their cumulative probability, and the next $n$ best sequences are passed to the next decoder state. $n$ is often called the width or beam size. When increasing the beam size, the benefit of having more proposals is weighed against the cost of evaluating their correctness (such as compilation, running tests, and executing a process to confirm the vulnerability is repaired).

\textbf{Definition} The \emph{Interpretation beam} is the number of programs that can be created from a given input function given a \emph{token context diff} change specification. The interpretation beam is specific to our change representation. For example, \autoref{fig:context} shows a case where a 3 token context size has only 1 possible application, while the 2 token context can be interpreted in 3 different ways. Hence, the interpretation beam is 3 for context size 2, and 1 for context size 3.

\textbf{Definition} The \emph{VRepair beam} is the combination of the neural beam and the interpretation beam, it is the cartesian product of all programs that can be created from both the Neural beam and the Interpretation beam.

Once VRepair outputs a patch, the patched function is meant to be verified by a human, a test suit, or a static analyzer, depending on the software process. For example, if the vulnerability was found by a static analyzer, the patched program can be verified again by the same static analyzer to confirm that the vulnerability has been fixed by VRepair. As this evaluation may consume time and resources, we evaluate the output of the VRepair system by setting the limit on \emph{VRepair beam} which represents the number of programs proposed by VRepair for evaluation.

\subsection{Neural Network Architecture}
\label{sec:architecture}

\begin{figure}
    \centering
    \includegraphics[width=0.45\textwidth]{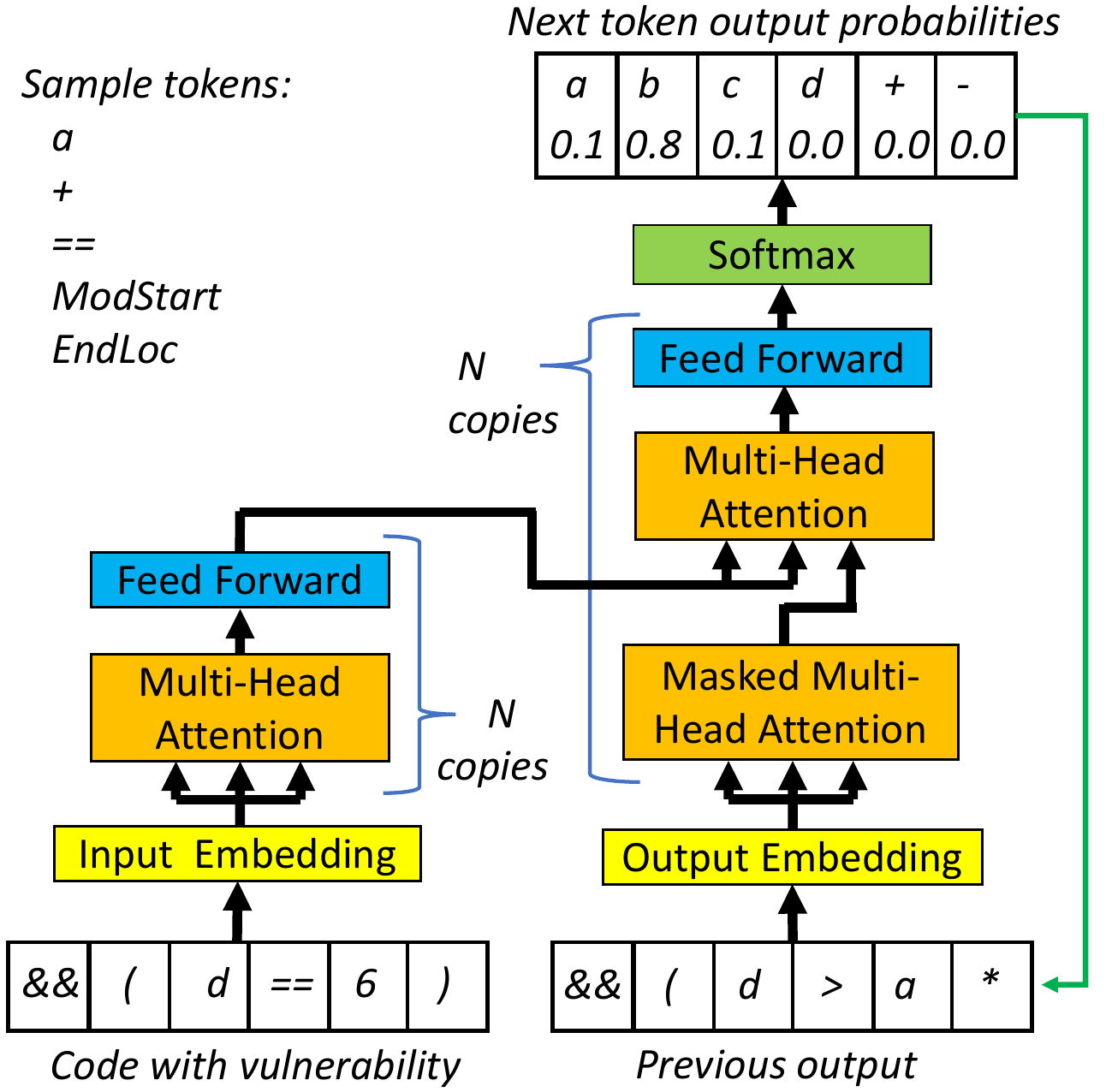}
    \caption{VRepair Neural Architecture based on Transformers}
    \label{fig:transformer}
\end{figure}

VRepair uses the Transformer architecture \cite{vaswani2017attention} as sketched in \autoref{fig:transformer}. The Transformer for VRepair learns to repair vulnerabilities by first receiving as input the code with a vulnerability encoded in our data representation (\autoref{subsec:data_representation}). Multiple copies of multi-head attention layers learn hidden representations of the input data. These representations are then used by a second set of multi-head attention layers to produce a table of probabilities for the most likely token to output. In addition, we use a copy mechanism that trains an alternate neural network to learn to copy input tokens to the output in order to reduce the vocabulary size required \cite{gu2016incorporating,chen2019sequencer}.

The first token to output is based solely on the hidden representations the model has learned to produce from the input code. As tokens are output they are available as input to the model so it can adjust the next token probabilities correctly. For example, in \autoref{fig:transformer}, after the sequence of tokens '\texttt{\&\& ( d > a *}' has been output, the model predicts that the next token should be '\texttt{b}' with a probability of 0.8. 

\emph{Libraries} VRepair is implemented in Python using state-of-the-art tools. We download all GitHub events using GH Archive \cite{GHArchive}. For processing the source code we use GCC and Clang. Once the source code is processed, OpenNMT-py is used to train the core Transformer model \cite{2017opennmt}.

\emph{Hyperparameters} Hyperparameters define the particular model and dataset configuration for VRepair and our primary hyperparameters are presented in \autoref{tab:params}. The context size is a hyperparameter specific to VRepair and is discussed in more detail in \autoref{subsec:data_representation}. We include the most common 80\% of CWE IDs among our 2000 word vocabulary. For both learning rate and the hidden and embedding size, we do a hyperparameter search for finding the best model among the values specified in \autoref{tab:params}. The learning rate decay is 0.9, meaning that our model decays the learning rate by 0.9 for every \num{10000} training steps, starting from step \num{50000}.

\emph{Beam Width}
An important parameter during inference is the beam width. Since the majority of papers on code generation use a value of 50 for neural beam size \cite{chen2019sequencer,tufano2019empirical,ahmed2018compilation}, we also select this number. Recall that the interpretation beam combines with the neural beam (see \autoref{sec:inference}), which may increase the number of proposals. Hence, we also set the \emph{VRepair beam} width to 50 (50 possible predictions per input). Our ablation study in \autoref{sec:ablation} provides data demonstrating that a beam width of 50 produces more accurate results than smaller beam widths we tested.

\begin{table}[]
\centering
\caption{Training Hyperparameters in VRepair.}
\label{tab:params}
\resizebox{.8\linewidth}{!}{\begin{tabular}{ll}
\hline
\multicolumn{1}{c}{Hyperparameter}                          & Value                                       \\ \hline
Context size                                                & 3                                           \\
Batch size                                                  & 32                                          \\
Vocabulary size                                             & [2000, 5000, 10000]                         \\
Layers (N copies of attention)                              & 6                                           \\
Optmizer                                                    & Adam optmizer \cite{kingma2014adam}         \\
Learning rate                                               & [$5e^{-4}, 1e^{-4}, 5e^{-5}$]               \\
Hidden and embedding size                                   & [256, 512, 1024]                            \\
Feed forward size                                           & 2048                                        \\
Dropout \cite{srivastava2014dropout}                        & 0.1                                         \\
Label smoothing \cite{szegedy2016rethinking}                & 0.1                                         \\
Attention heads                                             & 8                                           \\
Learning rate decay                                         & 0.9                                         \\ \hline
\end{tabular}}
\end{table}

The scripts and data that we use are available at \url{https://github.com/SteveKommrusch/VRepair}. 

\subsection{Usage of VRepair}

As indicated by \autoref{fig:workflow}, VRepair is intended to be used for proposing vulnerability fixes within an environment that provides vulnerability detection. Recall that vulnerability detection is not in the scope of VRepair per se, tools such as Infer can be used to statically analyze code and indicate a suspicious line containing a vulnerability (see \autoref{subsec:data_representation}). Once a patch is generated, the expected use case is to recompile the function with the proposed vulnerability fix, to pass a functional test suite if such a test suite exists, and then to pass the vulnerability checker under consideration (such as Infer). As the last step, having the patch reviewed by a human is likely to be done in a practical setting. The human review would occur after a patch has been shown to compile and pass the test suite and vulnerability oracle, in order to save human effort. It is also possible to integrate VRepair in continuous integration, based on the blueprint architecture of R-Hero \cite{Baudry21Repair}.

%%%%%%%%%%%%%%%%%%%%%%%%%%%%%%%%%%%%%%%%%%
%%%%%%%%%%%%%%%%%%%%%%%%%%%%%%%%%%%%%%%%%
\section{Experimental Protocol}
\label{sec:protocol}

In this section, we describe our methodology for evaluating our approach by defining 4 research questions and how we propose to answer them.

\subsection{Research Questions}

\begin{enumerate}
\item RQ1: What is the accuracy of only source or only target domain training on the vulnerability fixing task?
\item RQ2: What is the accuracy of transfer learning with source and target domain training on the vulnerability fixing task?
\item RQ3: What is the accuracy of transfer learning compared to denoising pre-training?
\item RQ4: How do different data split strategies impact the accuracy of models trained with transfer learning and target domain training?
\end{enumerate}

RQ1 will explore the performance of the neural model when only trained with the small dataset available from the target domain (vulnerability fixing), or only trained with a larger dataset from the bug fixing source domain; both being called `single domain training'. After exploring single domain training, RQ2 will study the effectiveness of using transfer learning to mitigate the issue of the small vulnerability dataset size demonstrated in RQ1. For RQ2, we study whether a model trained on the source domain (bug fixing) and then tuned with the target domain (vulnerability repair), produces a better result than either source or target domain training alone. Once we have explored transfer learning, in RQ3, we will investigate the possibility of using an unsupervised denoising pre-training technique to alleviate the small dataset problem. In RQ4, we would like to understand how transfer learning's measurement is affected when we split the evaluation data with different strategies.

\subsection{Datasets} \label{subsec:dataset}

We create a bug fixing dataset for the purpose of the experiment (see \autoref{sec:bugfixcorpus}). For vulnerabilities, we use two existing vulnerability fix datasets from the literature, called Big-Vul \cite{fan2020ac} and CVEfixes \cite{bhandari2021cvefixes} that both consist of confirmed vulnerabilities with CVE IDs.

\subsubsection{Bug Fix Corpus}
\label{sec:bugfixcorpus}

We create a bug fix corpus by mining the GitHub development platform. 
We follow the procedures in related works \cite{tufano2019empirical, lutellier2019encore} and collect our bug fixing dataset by filtering GitHub commits to C code projects based on keywords such as 'bug' or 'vulnerability' in the commit message. Filtering commits based on commit messages can be imprecise and generate false positives. However, this imprecision has minimal effect for our source domain dataset - any code change will likely help train the model on how to propose code patches.

We download 892 million GitHub events from the GH Archive data \cite{GHArchive} which happened between 2017-01-01 and 2018-12-31. These events have been triggered by a Github issue creation, an opened pull request, and other development activities. In our case, we focus on push events, which are triggered when at least one commit is pushed to a repository branch. In total there were 478 million push events between 2017-01-01 and 2018-12-31.

% only bug fix commits
Next, we filter bug fix commits as follows. Per the related work \cite{tufano2019empirical, lutellier2019encore}, we adopt a keyword-based heuristic: if the commit message contains keywords (\textit{fix} OR \textit{solve} OR \textit{repair}) AND (\textit{bug} OR \textit{issue} OR \textit{problem} OR \textit{error} OR \textit{fault} OR \textit{vulnerability}), we consider it a bug fix commit and add it to our corpus. In total, we have analyzed 729 million (\num{728916054}) commits and selected 21 million (\num{20568128}) commits identified as bug fix commits. This is a dataset size that goes beyond prior work by Tufano \etal \cite{tufano2019empirical}, who uses 10 million (\num{10056052}) bug-fixing commits.

% focus on C
In our experiment, we focus on C code as the target programming language for automatic repair. Therefore we further filter the bug fix commits based on the file extension and we remove commits that did not fix any file that ends with '.c', resulting in \num{910000} buggy C commits.

For each changed file in each commit, we compare all functions before and after the change to extract functions that changed; we call these function pairs. To identify these function pair changes, we use the GNU compiler preprocessor to remove all comments, and we extract functions with the same function signature in order to compare them. Then, we used Clang \cite{Clang} to parse and tokenize the function's source code. Within a '.c' file we ensure that only full functions (not prototypes) are considered. 

In the end, we obtain \num{1838740} function-level changes, reduced to \num{655741} after removing duplicate functions. The large number of duplicates can be explained by code clones, where the same function is implemented in multiple GitHub projects. As detailed in Section~\ref{subsec:data_representation}, our preferred context size is 3 tokens; when duplicate functions plus change specifications are removed using this representation, we have \num{650499} unique function plus specification samples.

For all of our experiments, we limit the input length of functions to 1000 tokens and the token context diff output to 100 tokens in order to limit the memory needs of the model. A 100 token output limit has been seen to produce quality results for machine learning on code \cite{tufano2019empirical}. For all research questions, we partition this dataset into $B_{train}$ (for model training, \num{534858} samples) and $B_{val}$ (for model validation, \num{10000} samples).

\subsubsection{Vulnerability Fix Corpus}

We use two existing datasets called Big-Vul \cite{fan2020ac} and CVEfixes \cite{bhandari2021cvefixes} for tuning the model trained on the bug fixing examples. The Big-Vul dataset has been created by crawling CVE databases and extracting vulnerability related information such as CWE ID and CVE ID. Then, depending on the project, the authors developed distinct crawlers for each project's pages to obtain the git commit link that fixes the vulnerability. In total, Big-Vul contains \num{3754} different vulnerabilities across \num{348} projects categorized into 91 different CWE IDs, with a time frame spanning from 2002 to 2019.

The CVEfixes dataset is collected in a way similar to the Big-Vul dataset. This dataset contains \num{5365} vulnerabilities across \num{1754} projects categorized into 180 different CWE IDs, with a time frame spanning from 1999 to 2021. By having two datasets collected in two independent papers, we can have higher confidence about the generalizability of our conclusions. All the research questions will be done on both vulnerability fix datasets.

\subsection{Methodology for RQ1} \label{subsec:method_rq1}

\begin{table}
\begin{minipage}[t]{.48\linewidth}\centering
\resizebox{\linewidth}{!}{\begin{tabular}{cc}
\hline
        & Examples in \\ 
CWE ID  & train/valid/test \\ \hline
CWE-119 & 698/108/187                  \\
CWE-20  & 152/25/51                    \\
CWE-125 & 164/15/45                    \\
CWE-264 & 129/16/32                    \\
CWE-399 & 95/19/29                     \\
CWE-200 & 103/19/28                    \\
CWE-476 & 87/12/26                     \\
CWE-284 & 66/10/26                     \\
CWE-189 & 58/8/26                      \\
CWE-362 & 76/2/26                      \\ \hline
\end{tabular}}
\caption{Number of top-10 CWE examples from the $\textnormal{Big-Vul}_{test}^{rand}$ in $\textnormal{Big-Vul}_{train}^{rand}$, $\textnormal{Big-Vul}_{val}^{rand}$ and $\textnormal{Big-Vul}_{test}^{rand}$}
\label{tab:rq1_big_vul_cwe}
\end{minipage}
\hfill
\begin{minipage}[t]{.48\linewidth}\centering
\resizebox{\linewidth}{!}{\begin{tabular}{ccc}
\hline
        & Examples in \\ 
CWE ID  & train/valid/test \\ \hline
CWE-119 & 322/37/102                   \\
CWE-20  & 216/26/55                    \\
CWE-125 & 244/34/55                    \\
CWE-476 & 118/23/39                    \\
CWE-362 & 110/9/30                     \\
CWE-190 & 98/17/29                     \\
CWE-399 & 75/8/26                      \\
CWE-264 & 105/18/26                    \\
CWE-787 & 97/15/24                     \\
CWE-200 & 113/12/22                    \\ \hline
\end{tabular}}
\caption{Number of top-10 CWE examples from the $\textnormal{CVEfixes}_{test}^{rand}$ in $\textnormal{CVEfixes}_{train}^{rand}$, $\textnormal{CVEfixes}_{val}^{rand}$ and $\textnormal{CVEfixesl}_{test}^{rand}$
}
\label{tab:rq1_cve_fixes_cwe}
\end{minipage}
\end{table}

In this research question, we would like to understand the performance of a model which is trained with source domain only or target domain only. For our source domain dataset, we train on $B_{train}$ and validate with $B_{val}$, as detailed in \autoref{sec:bugfixcorpus}. Next, we randomly divide the vulnerable and fixed function pairs from Big-Vul into training $\textnormal{Big-Vul}_{train}^{rand}$, validation $\textnormal{Big-Vul}_{val}^{rand}$ and testing $\textnormal{Big-Vul}_{test}^{rand}$ sets, with \num{2226} (70\%), \num{318} (10\%) and \num{636} (20\%) examples in each respective set. Similarity, we also randomly divide the vulnerable and fixed function pairs from CVEfixes into training $\textnormal{CVEfixes}_{train}^{rand}$, validation $\textnormal{CVEfixes}_{val}^{rand}$ and testing $\textnormal{CVEfixes}_{test}^{rand}$ sets, with \num{2383} (70\%), \num{340} (10\%) and \num{681} (20\%) examples in each respective set. $\textnormal{Big-Vul}_{test}^{rand}$ and $\textnormal{CVEfixes}_{test}^{rand}$ are used to evaluate the models trained with source and target domain training. For all of the data splits in each dataset, we make sure that all of the examples are mutually exclusive, as recommended by Allamanis \cite{allamanis2019adverse}. The number of examples for the top-10 CWE values from $\textnormal{Big-Vul}_{test}^{rand}$ and $\textnormal{CVEfixes}_{test}^{rand}$ are given in \autoref{tab:rq1_big_vul_cwe} and \autoref{tab:rq1_cve_fixes_cwe}.

For the model with only source domain training, we train on $B_{train}$ and apply early stopping with $B_{val}$. The model is then evaluated by using a VRepair beam size of 50 on each example in $\textnormal{Big-Vul}_{test}^{rand}$ and $\textnormal{CVEfixes}_{test}^{rand}$, and the sequence accuracy is used as the performance metric. The sequence accuracy is 1 if any prediction sequence among the 50 outputs matches the ground truth sequence, and it is 0 otherwise. We compute the average test sequence accuracy over all examples in $\textnormal{Big-Vul}_{test}^{rand}$ and $\textnormal{CVEfixes}_{test}^{rand}$.

For the model with only target domain training, we train on $\textnormal{Big-Vul}_{train}^{rand}$ (or $\textnormal{CVEfixes}_{train}^{rand}$) and apply early stopping on $\textnormal{Big-Vul}_{val}^{rand}$ (or $\textnormal{CVEfixes}_{val}^{rand}$). The model is then evaluated by using VRepair beam size 50 on each example in $\textnormal{Big-Vul}_{test}^{rand}$ (or $\textnormal{CVEfixes}_{test}^{rand}$), and the predictions are used to calculate the sequence accuracy. We hypothesize that the test sequence accuracy with source domain training will be lower than target domain training. This is because in source domain training, the training dataset is from a different domain, \ie bug fixes. But the result will show if a model trained on a large number of bug fixes can perform as well as a model trained on a small number of vulnerability repairs on the target task of vulnerability repair.

\subsection{Methodology for RQ2} \label{subsec:method_rq2}

The state-of-the-art vulnerability fixing models are usually trained on a relatively small vulnerability fixing dataset \cite{chi2020seqtrans}, or generated from synthesized code examples \cite{harer2018learning}. Based on prior studies showing the effectiveness of large datasets for machine learning \cite{sun2017revisiting}, we hypothesize that it is hard for a deep learning model to generalize well on such a small dataset. In this research question, we compare the performance between the transfer learning model and the model only trained on the small vulnerability fix dataset.

We first train the models with source domain training, \ie we train them on $B_{train}$ and apply early stopping on $B_{val}$. The models from source domain training are used in the target domain training phase. We continue training the models using  $\textnormal{Big-Vul}_{train}^{rand}$ (or $\textnormal{CVEfixes}_{train}^{rand}$) and the new model is selected based on $\textnormal{Big-Vul}_{val}^{rand}$ (or $\textnormal{CVEfixes}_{val}^{rand}$) with early stopping. During the target domain training phase, per the standard practice of lowering the learning rate when learning in the target domain \cite{shin2016deep}, we use a learning rate that is one tenth of the learning rate used in the source domain training phase. Finally, the final model is evaluated with $\textnormal{Big-Vul}_{test}^{rand}$ (or $\textnormal{CVEfixes}_{test}^{rand}$) by using VRepair beam size 50 on each example, which we will use to calculate the sequence accuracy.

\subsection{Methodology for RQ3} \label{subsec:method_rq3}

We compare our source domain training with the state of the art pre-training technique of PLBART \cite{ahmad2021unified}, which is based on denoising. \minor{This means  we first train on an unsupervised  task (the denoising task). The trained model is then fine-tuned with target domain training as we do for VRepair. We hypothesize that initial training on a related task is more helpful than initial training on a generic unsupervised task.} Specifically, we select full functions from our bug fix corpus and apply PLBART's noise function techniques for token masking, token deletion, and token infilling. This results in artificial 'buggy' functions whose target repair is the original function. As per PLBART (and the original BART paper addressing natural language noise functions \cite{lewis2019bart}): token masking replaces tokens with a special token \texttt{<MASK>} in the 'buggy' function; token deletion removes tokens; and token infilling replaces multiple consecutive tokens with a single \texttt{<MASK>} token. As in PLBART, we sample from a Poisson distribution to determine the length for the token infilling noise function.

Following this methodology, we generate the pre-training dataset from the bug fix corpus divided into $\textnormal{Pre}_{train}$ (\num{728730} samples) and $\textnormal{Pre}_{val}$ (\num{10000} samples). For the pre-trained model, similar to \autoref{subsec:method_rq2}, we first train the models on $\textnormal{Pre}_{train}$ and apply early stopping on $\textnormal{Pre}_{val}$. We continue training the models using $\textnormal{Big-Vul}_{train}^{rand}$ (or $\textnormal{CVEfixes}_{train}^{rand}$) and the new model is selected based on $\textnormal{Big-Vul}_{val}^{rand}$ (or $\textnormal{CVEfixes}_{val}^{rand}$) with early stopping. Finally, the final model is evaluated with $\textnormal{Big-Vul}_{test}^{rand}$ (or $\textnormal{CVEfixes}_{test}^{rand}$), and compared against the transfer learning models trained in \autoref{subsec:method_rq2}.

% Details of our masking are given 
% in src/prespecial.pl
% \revision{Our noise injection algorithm is as follows: process the original function sample one token at a time; with 5\% probability begin a noise function on a given token. The noise function is selected uniformly from masking, deletion, and infilling. Continue the chosen noise function for a length selected from a Poisson distribution with $\lambda=3$, creating the 'buggy' input as well as the 3-token context target output as tokens are processed. When a change completes, gather 3 more tokens without noise to cleanly complete the context of the change. After the change context is complete, continue processing the function one token at a time with 5\% probability of beginning a new random noise function on a given token. In this way, multiple different noise functions may be applied within a given function, much like change patterns we observe in actual code. {Is the above already too much detail, or should we keep it? Should we include the short function in Steve's email which shows all 3 noise types in an example?}}

\subsection{Methodology for RQ4}
\label{subsec:method_rq4}

\begin{table}
\begin{minipage}[t]{.48\linewidth}\centering
\resizebox{\linewidth}{!}{\begin{tabular}{cc}
\hline
        & Examples in \\ 
CWE ID  & train/valid/test \\ \hline
CWE-119 & 835/41/117                   \\
CWE-125 & 71/78/75                     \\
CWE-416 & 48/19/47                     \\
CWE-476 & 64/15/46                     \\
CWE-190 & 48/3/43                      \\
CWE-787 & 14/13/27                     \\
CWE-20  & 165/37/26                    \\
CWE-200 & 103/21/26                    \\
CWE-362 & 68/11/25                     \\
CWE-415 & 9/0/13                       \\ \hline
\end{tabular}}
\caption{Number of top-10 CWE examples from the $\textnormal{Big-Vul}_{test}^{year}$ in $\textnormal{Big-Vul}_{train}^{year}$, $\textnormal{Big-Vul}_{val}^{year}$ and $\textnormal{Big-Vul}_{test}^{year}$}
\label{tab:rq4_big_vul_cwe}
\end{minipage}
\hfill
\begin{minipage}[t]{.48\linewidth}\centering
\resizebox{\linewidth}{!}{\begin{tabular}{ccc}
\hline
        & Examples in \\ 
CWE ID  & train/valid/test \\ \hline
CWE-125 & 172/35/126                   \\
CWE-20  & 185/11/101                   \\
CWE-787 & 37/44/55                     \\
CWE-476 & 107/25/48                    \\
CWE-119 & 396/20/45                    \\
CWE-416 & 48/24/30                     \\
CWE-190 & 92/24/28                     \\
CWE-362 & 117/11/21                    \\
CWE-200 & 123/14/10                    \\
CWE-415 & 12/17/9                      \\ \hline
\end{tabular}}
\caption{Number of top-10 CWE examples from the $\textnormal{CVEfixes}_{test}^{year}$ in $\textnormal{CVEfixes}_{train}^{year}$, $\textnormal{CVEfixes}_{val}^{year}$ and $\textnormal{CVEfixesl}_{test}^{year}$}
\label{tab:rq4_cve_fixes_cwe}
\end{minipage}
\end{table}

In this experiment, we wish to study how the performance of a vulnerability fixing model varies with different data split strategies compared to only target domain training. For this, we divide the Big-Vul and CVEfixes dataset using two strategies:
1) random, as done previously in \autoref{subsec:method_rq1}, \autoref{subsec:method_rq2}, and \autoref{subsec:method_rq3};
2) time-based;

\emph{Time-based splitting} First, we sort the Big-Vul dataset based on CVE publication dates. Recall that Big-Vul contains vulnerabilities collected from 2002 to 2019. We create a testing set $\textnormal{Big-Vul}_{test}^{year}$ of 603 data points, containing all buggy and fixed function pairs with a published date between 2018-01-01 to 2019-12-31. The validation set $\textnormal{Big-Vul}_{val}^{year}$ (302 examples) contains all buggy and fixed function pairs with a published date between 2017-06-01 to 2017-12-31, and the rest are in the training set $\textnormal{Big-Vul}_{train}^{year}$ (2272 examples).

Similarly, for the CVEfixes dataset, we create a testing set $\textnormal{CVEfixes}_{test}^{year}$ of 794 data points, containing all buggy and fixed function pairs with a published date between 2019-06-01 to 2021-06-09. The validation set $\textnormal{CVEfixes}_{val}^{year}$ (324 examples) contains all buggy and fixed function pairs with a published date between 2018-06-01 to 2019-06-01, and the rest are in the training set $\textnormal{CVEfixes}_{train}^{year}$ (2286 examples). The dates are chosen so that we have roughly 70\% data in the training set, 10\% data in the validation set and 20\% data in the test set. This data split strategy simulates a vulnerability fixing system that is trained on past vulnerability fixes, and that is used to repair vulnerabilities in the future. The number of examples of the top-10 CWE from $\textnormal{Big-Vul}_{test}^{year}$ and $\textnormal{CVEfixes}_{test}^{year}$ are given in \autoref{tab:rq4_big_vul_cwe} and \autoref{tab:rq4_cve_fixes_cwe}.

For all strategies, we train two different models with source+target domain training (transfer learning) and only target domain training, following the same protocol in \autoref{subsec:method_rq1}, \autoref{subsec:method_rq2}, and \autoref{subsec:method_rq3}, but with different data splits. We report the test sequence accuracy on all data splits.

\section{Experimental Results}
We now present the results of our large scale empirical evaluation of VRepair, per the experimental protocol presented in \autoref{sec:protocol}.

\subsection{Results for RQ1}
\label{subsec:result_rq1}

We study the test sequence accuracy of models trained only with source or only with target domain training. Given our datasets, source domain training means training the model only with bug fix examples from $B_{train}$, while target domain training uses only $\textnormal{Big-Vul}_{train}^{rand}$ (or $\textnormal{CVEfixes}_{train}^{rand}$). Both models are evaluated on the vulnerability fixing examples in $\textnormal{Big-Vul}_{test}^{rand}$ (or $\textnormal{CVEfixes}_{test}^{rand}$). \autoref{tab:rq1_big_vul} gives the results on $\textnormal{Big-Vul}_{test}^{rand}$. In the first column we list the top-10 most common CWE IDs in $\textnormal{Big-Vul}_{test}^{rand}$. The second column shows the performance of the model only trained with source domain training. The third column shows the performance of the model only trained with target domain training. The last row of the table presents the test sequence accuracy on the whole$\textnormal{Big-Vul}_{test}^{rand}$. The result for $\textnormal{CVEfixes}_{test}^{rand}$ is in \autoref{tab:rq1_cvefixes}.

\begin{table}[]
\caption{RQ1: Test sequence accuracy on $\textnormal{Big-Vul}_{test}^{rand}$ with source/target domain training, we also present the accuracy on the top-10 most common CWE IDs in $\textnormal{Big-Vul}_{test}^{rand}$. The absolute numbers represent \# of C functions in test dataset.}
\label{tab:rq1_big_vul}
\resizebox{\linewidth}{!}{\begin{tabular}{ccc}
\hline
CWE ID  & Source domain training & Target domain training \\ \hline
CWE-119 & 12.30\% (23/187)       & 11.76\% (22/187)       \\
CWE-20  & 19.61\% (10/51)        & 11.76\% (6/51)         \\
CWE-125 & 17.78\% (8/45)         & 8.89\% (4/45)          \\
CWE-264 & 6.25\% (2/32)          & 6.25\% (2/32)          \\
CWE-399 & 24.14\% (7/29)         & 0.00\% (0/29)          \\
CWE-200 & 32.14\% (9/28)         & 0.00\% (0/28)          \\
CWE-476 & 19.23\% (5/26)         & 7.69\% (2/26)          \\
CWE-284 & 61.54\% (16/26)        & 23.08\% (6/26)         \\
CWE-189 & 19.23\% (5/26)         & 3.85\% (1/26)          \\
CWE-362 & 23.08\% (6/26)         & 0.00\% (0/26)          \\
All     & 18.24\% (116/636)      & 7.86\% (50/636)        \\ \hline
\end{tabular}}
\end{table}

\begin{table}[]
\caption{RQ1: Test sequence accuracy on $\textnormal{CVEfixes}_{test}^{rand}$ with source/target domain training.}
\label{tab:rq1_cvefixes}
\resizebox{\linewidth}{!}{\begin{tabular}{ccc}
\hline
CWE ID  & Source domain training & Target domain training \\ \hline
CWE-119 & 8.82\% (9/102)         & 12.75\% (13/102)         \\
CWE-20  & 18.18\% (10/55)        & 12.73\% (7/55)        \\
CWE-125 & 18.18\% (10/55)        & 7.27\% (4/55)          \\
CWE-476 & 20.51\% (8/39)         & 12.82\% (5/39)         \\
CWE-362 & 3.33\% (1/30)          & 0.00\% (0/30)          \\
CWE-190 & 24.14\% (7/29)         & 34.48\% (10/29)         \\
CWE-399 & 19.23\% (5/26)         & 0.00\% (0/26)          \\
CWE-264 & 3.85\% (1/26)          & 11.54\% (3/26)          \\
CWE-787 & 8.33\% (2/24)          & 0\% (0/24)             \\
CWE-200 & 36.36\% (8/22)         & 4.55\% (1/22)         \\
All     & 15.98\% (109/682)      & 11.29\% (77/682)       \\ \hline
\end{tabular}}
\end{table}

% source domain training and target domain training
Even when the model is trained on the different domain of bug fixing, the model still achieves a \minor{18.24\% accuracy on $\textnormal{Big-Vul}_{test}^{rand}$ and 15.98\% accuracy} on $\textnormal{CVEfixes}_{test}^{rand}$. This is better than the models that are trained with only target domain training, \ie training on a small vulnerability fix dataset, that has a performance of \minor{7.86\% on $\textnormal{Big-Vul}_{test}^{rand}$ and 11.29\%} on $\textnormal{CVEfixes}_{test}^{rand}$. The result shows that training on a small dataset indeed is ineffective and even training on a bug fix corpus, which is a different domain but has a bigger dataset size, will increase the performance.

% comparison over CWE
If we compare the results across the top-10 most common CWE IDs in both $\textnormal{Big-Vul}_{test}^{rand}$ and $\textnormal{CVEfixes}_{test}^{rand}$, we see that the source domain trained models outperform the target domain trained models over almost all CWE categories. \minor{In $\textnormal{CVEfixes}_{test}^{rand}$, there are some exceptions such as CWE-119, CWE-190 and CWE-264, where the target domain trained model slightly surpassed the performance of the source domain trained model. However the difference is small and the overall performance of source domain trained models dominates the target domain trained models.}

% listing
\minor{In \autoref{lst:CVE-2018-11375} we show an example of a vulnerability fix that was correctly predicted by the model trained on the target domain only. The vulnerability is CVE-2018-11375 with type CWE-125 (Out-of-bounds Read) from the Radare2 project. CVE-2018-11375 can cause a denial of service attack via a crafted binary file\footnote{\url{https://nvd.nist.gov/vuln/detail/CVE-2018-11375}}. The vulnerability is fixed by checking the length of variable \textit{len}, which is correctly predicted by the VRepair model.}

\noindent\begin{minipage}{\linewidth}
\centering
\begin{lstlisting}[language=diff,columns=flexible, frame=single, basicstyle=\ttfamily\scriptsize, label={lst:CVE-2018-11375}, caption={CVE-2018-11375 is correctly predicted by the model trained with target domain only.}, captionpos=b, breaklines=true]
}

INST_HANDLER (lds) {	// LDS Rd, k
+   if (len < 4) {
+       return;
+   }
 	int d = ((buf[0] >> 4) & 0xf) | ((buf[1] & 0x1) << 4);
 	int k = (buf[3] << 8) | buf[2];
 	op->ptr = k;
\end{lstlisting}
\end{minipage}

\begin{mdframed}[nobreak=true]
Answer to RQ1: Training a VRepair Transformer on a small vulnerability fix dataset achieves accuracies of \minor{7.86\% on Big-Vul and 11.29\% on CVEfixes}. Surprisingly, by training only on bug fixes, the same neural network achieves better accuracies of \minor{18.24\% on Big-Vul and 15.98\% on CVEfixes}, which shows the ineffectiveness of just training on a small vulnerability dataset.
\end{mdframed}

\subsection{Results for RQ2}
\label{subsec:result_rq2}

\begin{table}[]
\caption{RQ2: Test sequence accuracy on $\textnormal{Big-Vul}_{test}^{rand}$ with transfer learning. The 'Improvement over source domain training' and 'Improvement over target domain training' columns show the percentage and numerical improvement compared to the result in \autoref{tab:rq1_big_vul}. Transfer learning achieved better overall performance and on all CWE IDs.}
\label{tab:rq2_big_vul}
\resizebox{\linewidth}{!}{\begin{tabular}{cccc}
\hline
CWE ID  & Transfer learning & \thead{Improvement over \\ source domain training} &  \thead{Improvement over \\ target domain training} \\ \hline
CWE-119 & 17.65\% (33/187)  & +5.35\%/+10                                        & +5.89\%/+11 \\
CWE-20  & 19.61\% (10/51)   & +0.00\%/+0                                         & +7.85\%/+4   \\
CWE-125 & 17.78\% (8/45)    & +0.00\%/+0                                         & +8.89\%/+4  \\
CWE-264 & 6.25\% (2/32)     & +0.00\%/+0                                         & +0.00\%/+0   \\
CWE-399 & 34.48\% (10/29)   & +10.34\%/+3                                        & +34.48\%/+10  \\
CWE-200 & 35.71\% (10/28)   & +3.57\%/+1                                         & +35.71\%/+10  \\
CWE-476 & 26.92\% (7/26)    & +7.69\%/+2                                         & +19.23\%/+5  \\
CWE-284 & 65.38\% (17/26)   & +3.84\%/+1                                         & +42.30\%/+11  \\
CWE-189 & 19.23\% (5/26)    & +0.00\%/+0                                         & +15.38\%/+4  \\
CWE-362 & 30.77\% (8/26)    & +7.69\%/+2                                         & +30.77\%/+8  \\
All     & 21.86\% (139/636) & +3.62\%/+23                                        & +14.00\%/+89 \\ \hline
\end{tabular}}
\end{table}

\begin{table}[]
\caption{RQ2: Test sequence accuracy on $\textnormal{CVEfixes}_{test}^{rand}$ with transfer learning. The 'Improvement over source domain training' and 'Improvement over target domain training' columns show the percentage and numerical improvement compared to the result in \autoref{tab:rq1_cvefixes}.}
\label{tab:rq2_cvefixes}
\resizebox{\linewidth}{!}{\begin{tabular}{cccc}
\hline
CWE ID  & Transfer learning & \thead{Improvement over \\ source domain training} &  \thead{Improvement over \\ target domain training} \\ \hline
CWE-119 & 22.55\% (23/102)  & +13.73\%/+14                                       & +9.80\%/+10 \\
CWE-20  & 27.27\% (15/55)   & +9.09\%/+5                                         & +14.54\%/+8   \\
CWE-125 & 16.36\% (9/55)    & -1.82\%/-1                                         & +9.09\%/+5  \\
CWE-476 & 33.33\% (13/39)   & +12.82\%/+5                                        & +20.51\%/+8   \\
CWE-362 & 13.33\% (4/30)    & +10.00\%/+3                                        & +13.33\%/+4  \\
CWE-190 & 27.59\% (8/29)    & +3.45\%/+1                                         & -6.89\%-2  \\
CWE-399 & 23.08\% (6/26)    & +3.85\%/+1                                         & +23.08\%/+6  \\
CWE-264 & 11.54\% (3/26)    & +7.69\%/+2                                         & +0.00\%/+0  \\
CWE-787 & 12.50\% (3/24)    & +4.17\%/+1                                         & +12.50\%/+3  \\
CWE-200 & 45.45\% (10/22)   & +9.09\%/+2                                         & +40.90\%/+9  \\
All     & 22.73\% (155/682) & +6.75\%/+46                                        & +11.44\%/+78 \\ \hline
\end{tabular}}
\end{table}

In RQ2, we study the impact of transfer learning, \ie we measure the performance of a model with target domain training applied on the best model trained with source domain training. The results are given in \autoref{tab:rq2_big_vul} and \autoref{tab:rq2_cvefixes}. The first column lists the top-10 most common CWE IDs. The second column presents the performance of the model trained with transfer learning, \ie taking the source domain trained model from \autoref{tab:rq1_big_vul} (or \autoref{tab:rq1_cvefixes}) and tuning it with target domain training. The third and fourth column gives the performance increase compared to the model with only source or target domain training. The last row of the table presents the test sequence accuracy on the whole $\textnormal{Big-Vul}_{test}^{rand}$ (or $\textnormal{CVEfixes}_{test}^{rand}$). 

% main finding
The main takeaway is that the transfer learning model achieves the highest test sequence accuracy on both $\textnormal{Big-Vul}_{test}^{rand}$ and $\textnormal{CVEfixes}_{test}^{rand}$ with an accuracy of \minor{21.86\% and 22.73\%} respectively. Notably, transfer learning is superior to just target domain training on the small vulnerability fix dataset. In addition, transfer learning improves the accuracy over \minor{almost} all CWE categories when compared to only source domain training. This shows that the knowledge learned from the bug fix task can indeed be kept and fine-tuned to repair software vulnerabilities and that the previously learned knowledge from source domain training is useful. The result confirms that the bug fixing task and the vulnerability fixing task have similarities and that the bug fixing task can be used to train a model from which knowledge can be transferred.

Now, we compare the results across the top-10 most common CWE IDs in $\textnormal{Big-Vul}_{test}^{rand}$ and $\textnormal{CVEfixes}_{test}^{rand}$. For all rows except for \minor{CWE-125 and CWE-190 in \autoref{tab:rq2_cvefixes}}, the CWE ID performance is better with the transfer learning model. This is explained by the fact that the transfer learning model prefers generalization (fixing more CWE types) over specialization. The same phenomenon has been observed in a compiler error fix system, where pre-training lowers the performance of some specific compiler error types \cite{yasunaga2020graph}. Overall, the best VRepair transfer learning model is able to fix vulnerability types that are both common and rare.

\minor{Finally, we discuss a vulnerability where the VRepair transfer learning model is able to predict the exact fix, but not the target domain trained model. Vulnerability CVE-2016-9754, labeled with type CWE-190 (Integer Overflow or Wraparound) is shown in \autoref{lst:CVE-2016-9754}. The vulnerability can allow users to gain privileges by writing to a certain file\footnote{\url{https://nvd.nist.gov/vuln/detail/CVE-2016-9754}}. The vulnerability patch fixes the assignment of the variable \textit{size}, which previously could overflow. VRepair with transfer learning successfully predicts the patch, but not the target domain trained model, showing that the source domain training phase helps VRepair to fix a more sophisticated vulnerability.}

% multi-line claim
\autoref{lst:CVE-2016-9754} is also a notable case where VRepair successfully predicts a multi line patch. The data representation we described in \autoref{subsec:data_representation} allows VRepair to have a concise output representing this multi line patch. The token context diff makes it easier for VRepair to handle multi-line patches, since the output is shorter than the full functions used in related work \cite{tufano2019empirical}.

\noindent\begin{minipage}{\linewidth}
\centering
\begin{lstlisting}[language=diff,columns=flexible, frame=single, basicstyle=\ttfamily\footnotesize, label={lst:CVE-2016-9754}, caption={CVE-2016-9754 is correctly predicted by VRepair. It is an example of a multi line vulnerability fix, and the target domain trained model failed to predict the fix.}, captionpos=b, breaklines=true]
    !cpumask_test_cpu(cpu_id, buffer->cpumask))
    return size;
 
-size = DIV_ROUND_UP(size, BUF_PAGE_SIZE);
-size *= BUF_PAGE_SIZE;
+nr_pages = DIV_ROUND_UP(size, BUF_PAGE_SIZE);
 
 /* we need a minimum of two pages */
-if (size < BUF_PAGE_SIZE * 2)
-   size = BUF_PAGE_SIZE * 2;
+if (nr_pages < 2)
+   nr_pages = 2;
 
-nr_pages = DIV_ROUND_UP(size, BUF_PAGE_SIZE);
+size = nr_pages * BUF_PAGE_SIZE;
 
 /*
 * Don't succeed if resizing is disabled, as a reader might be
\end{lstlisting}
\end{minipage}

\begin{mdframed}[nobreak=true]
Answer to RQ2: By first learning from a large and generic bug fix dataset, and then tuning the model on the smaller vulnerability fix dataset, VRepair achieves better accuracies than just training on the small vulnerability fix dataset (\minor{21.86\% versus 7.86\% on $\textnormal{Big-Vul}_{test}^{rand}$ , and 22.73\% versus 11.29\% on $\textnormal{CVEfixes}_{test}^{rand}$}). Our experiment also shows that the VRepair transfer learning model is able to fix more rare vulnerability types.
\end{mdframed}

\subsection{Results for RQ3}
\label{subsec:result_rq3}

\begin{table}[h]
\caption{RQ3: Comparison between transfer learning and pre-training + target domain training on $\textnormal{Big-Vul}_{test}^{rand}$. The result shows the merit of first training on a related task, instead of generic pre-training.}
\label{tab:rq3_big_vul}
\resizebox{\linewidth}{!}{\begin{tabular}{ccc}
\hline
CWE ID  & Transfer learning & Pre-training + target domain training \\ \hline
CWE-119 & 17.65\% (33/187)  & 13.37\% (25/187)                      \\
CWE-20  & 19.61\% (10/51)   & 21.57\% (11/51)                       \\
CWE-125 & 17.78\% (8/45)    & 11.11\% (5/45)                        \\
CWE-264 & 6.25\% (2/32)     & 0.00\% (0/32)                         \\
CWE-399 & 34.48\% (10/29)   & 17.24\% (5/29)                        \\
CWE-200 & 35.71\% (10/28)   & 21.43\% (6/28)                        \\
CWE-476 & 26.92\% (7/26)    & 19.23\% (5/26)                        \\
CWE-284 & 65.38\% (17/26)   & 57.69\% (15/26)                       \\
CWE-189 & 19.23\% (5/26)    & 11.54\% (3/26)                        \\
CWE-362 & 30.77\% (8/26)    & 0.00\% (0/26)                         \\
All     & 21.86\% (139/636) & 13.36\% (85/636)                      \\ \hline
\end{tabular}}
\end{table}

\begin{table}[h]
\caption{RQ3: Comparison between transfer learning and pre-training + target domain training on $\textnormal{CVEfixes}_{test}^{rand}$.}
\label{tab:rq3_cvefixes}
\resizebox{\linewidth}{!}{\begin{tabular}{ccc}
\hline
CWE ID  & Transfer learning & Pre-training + target domain training \\ \hline
CWE-119 & 22.55\% (23/102)  & 10.78\% (11/102)                      \\
CWE-20  & 27.27\% (15/55)   & 16.36\% (9/55)                        \\
CWE-125 & 16.36\% (9/55)    & 9.09\% (5/55)                         \\
CWE-476 & 33.33\% (13/39)   & 10.26\% (4/39)                        \\
CWE-362 & 13.33\% (4/30)    & 3.33\% (1/30)                         \\
CWE-190 & 27.59\% (8/29)    & 20.69\% (6/29)                        \\
CWE-399 & 23.08\% (6/26)    & 15.38\% (4/26)                        \\
CWE-264 & 11.54\% (3/26)    & 3.85\% (1/26)                         \\
CWE-787 & 12.50\% (3/24)    & 20.83\% (5/24)                        \\
CWE-200 & 45.45\% (10/22)   & 31.82\% (7/22)                        \\
All     & 22.73\% (155/682) & 13.34\% (91/682)                      \\ \hline
\end{tabular}}
\end{table}

% core result about RQ3
RQ3 studies the effect of replacing the source domain training phase of transfer learning with denoising as pre-training in VRepair. As explained in \autoref{subsec:method_rq3}, we consider the state-of-the-art pre-training technique from PLBart \cite{ahmad2021unified}. In \autoref{tab:rq3_big_vul} and \autoref{tab:rq3_cvefixes}, we see that the transfer learning model dominates both test datasets, with \minor{21.86\% versus 13.36\%} on $\textnormal{Big-Vul}_{test}^{rand}$ and \minor{22.73\% versus 13.34\%} on $\textnormal{CVEfixes}_{test}^{rand}$. This means that first training on the bug fixing task is better than first training on a denoising task. But when comparing the same result against only training on the vulnerability fix dataset (target domain training) in \autoref{tab:rq1_big_vul} and \autoref{tab:rq1_cvefixes}, we can see that denoising pre-training does improve the result (\minor{7.86\% to 13.36\% on $\textnormal{Big-Vul}_{test}^{rand}$ and 11.29\% to 13.34\% on $\textnormal{CVEfixes}_{test}^{rand}$}). This shows that denoising pre-training is an alternative if collecting a large labeled source domain dataset is a hard task.

% discussion about pre-training size
From a qualitative perspective, denoising pre-training has the advantage of being unsupervised and therefore does not require collecting and curating a source domain dataset. Thanks to this property, CodeBERT \cite{feng2020codebert}, CuBERT \cite{kanade2020learning} and PLBART \cite{ahmad2021unified} all have millions of examples in the pre-training dataset. On the other hand, He, Girshick, and Dollár found that the performance of a pre-trained model scales poorly with the pre-training dataset size \cite{he2019rethinking}. This result shows that even when the size of the source domain dataset (\ie $B_{train}$ and $B_{val}$) is slightly smaller than the size of the pre-training dataset (\ie $Pre_{train}$ and $Pre_{val}$), transfer learning clearly outperforms denoising pre-training.

\begin{mdframed}[nobreak=true]
Answer to RQ3: In this experiment, transfer learning outperforms denoising pre-training and fine-tuning with datasets of similar size (\minor{21.86\% versus 13.36\% on $\textnormal{Big-Vul}_{test}^{rand}$, and 22.73\% versus 13.34\% on $\textnormal{CVEfixes}_{test}^{rand}$}). This result shows that the effort of collecting and curating a source domain dataset, arguably a tedious and consuming task, pays off with respect to performance. Overall, the specific source domain task of bug fixing is better than the generic task of denoising.
\end{mdframed}

\subsection{Results for RQ4}
\label{subsec:result_rq4}

\begin{table*}
\begin{minipage}{.24\linewidth}\centering
\resizebox{\linewidth}{!}{\begin{tabular}{ccc}
\hline
CWE ID  & Transfer learning & Target domain training \\ \hline
CWE-119 & 17.65\% (33/187)  & 11.76\% (22/187)        \\
CWE-20  & 19.61\% (10/51)   & 11.76\% (6/51)         \\
CWE-125 & 17.78\% (8/45)    & 8.89\% (4/45)          \\
CWE-264 & 6.25\% (2/32)     & 6.25\% (2/32)          \\
CWE-399 & 34.48\% (10/29)   & 0.00\% (0/29)           \\
CWE-200 & 35.71\% (10/28)   & 0.00\% (0/28)          \\
CWE-476 & 26.92\% (7/26)    & 7.69\% (2/26)          \\
CWE-284 & 65.38\% (17/26)   & 23.08\% (6/26)         \\
CWE-189 & 19.23\% (5/26)    & 3.85\% (1/26)          \\
CWE-362 & 30.77\% (8/26)    & 0.00\% (0/26)             \\
All     & 21.86\% (139/636) & 7.86\% (50/636)        \\ \hline
\end{tabular}}
\caption{Test sequence accuracy on testing data $\textnormal{Big-Vul}_{test}^{rand}$}
\label{tab:rq4_big_vul_rand}
\end{minipage}
\hfill
\begin{minipage}{.24\linewidth}\centering
\resizebox{\linewidth}{!}{\begin{tabular}{ccc}
\hline
CWE ID  & Transfer learning & Target domain training \\ \hline
CWE-119 & 20.51\% (24/117)  & 0.00\% (0/117)         \\
CWE-125 & 22.67\% (17/75)   & 12.00\% (9/75)         \\
CWE-416 & 25.53\% (12/47)   & 0.00\% (0/47)          \\
CWE-476 & 13.04\% (6/46)    & 0.00\% (0/46)          \\
CWE-190 & 13.95\% (6/43)    & 0.00\% (0/43)          \\
CWE-787 & 3.70\% (1/27)     & 0.00\% (0/27)          \\
CWE-20  & 15.38\% (4/26)    & 0.00\% (0/26)          \\
CWE-200 & 19.23\% (5/26)    & 0.00\% (0/26)          \\
CWE-362 & 12.00\% (3/25)    & 0.00\% (0/25)          \\
CWE-415 & 23.08\% (3/13)    & 0.00\% (0/13)          \\
All     & 19.24\% (116/603) & 1.49\% (9/603)         \\ \hline
\end{tabular}}
\caption{Test sequence accuracy on testing data $\textnormal{Big-Vul}_{test}^{year}$}
\label{tab:rq4_big_vul_year}
\end{minipage}
\hfill
\begin{minipage}{.24\linewidth}\centering
\resizebox{\linewidth}{!}{\begin{tabular}{ccc}
\hline
CWE ID  & Transfer learning & Target domain training \\ \hline
CWE-119 & 22.55\% (23/102)  & 12.75\% (13/102)       \\
CWE-20  & 27.27\% (15/55)   & 12.73\% (7/55)         \\
CWE-125 & 16.36\% (9/55)    & 7.27\% (4/55)          \\
CWE-476 & 33.33\% (13/39)   & 12.82\% (5/39)         \\
CWE-362 & 13.33\% (4/30)    & 0.00\% (0/30)          \\
CWE-190 & 27.59\% (8/29)    & 34.48\% (10/29)        \\
CWE-399 & 23.08\% (6/26)    & 0.00\% (0/26)          \\
CWE-264 & 11.54\% (3/26)    & 11.54\% (3/26)         \\
CWE-787 & 12.50\% (3/24)    & 0.00\% (0/24)          \\
CWE-200 & 45.45\% (10/22)   & 4.55\% (1/22)          \\
All     & 22.73\% (155/682) & 11.29\% (77/682)       \\ \hline
\end{tabular}}
\caption{Test sequence accuracy on testing data $\textnormal{CVEfixes}_{test}^{rand}$}
\label{tab:rq4_cvefixes_rand}
\end{minipage}
\hfill
\begin{minipage}{.24\linewidth}\centering
\resizebox{\linewidth}{!}{\begin{tabular}{ccc}
\hline
CWE ID  & Transfer learning & Target domain training  \\ \hline
CWE-125 & 20.63\% (26/126)  & 0.00\% (0/126)             \\
CWE-20  & 5.94\% (6/101)    & 0.00\% (0/101)             \\
CWE-787 & 1.82\% (1/55)     & 0.00\% (0/55)              \\
CWE-476 & 20.83\% (10/48)   & 0.00\% (0/48)              \\
CWE-119 & 13.33\% (6/45)    & 0.00\% (0/45)              \\
CWE-416 & 16.67\% (5/30)    & 0.00\% (0/30)              \\
CWE-190 & 42.86\% (12/28)   & 0.00\% (0/28)              \\
CWE-362 & 0.00\% (0/21)     & 0.00\% (0/21)              \\
CWE-200 & 0.00\% (0/10)     & 0.00\% (0/10)              \\
CWE-415 & 11.11\% (1/9)     & 0.00\% (0/9)               \\
All     & 16.23\% (129/795) & 0.13\% (1/795)             \\ \hline
\end{tabular}}
\caption{Test sequence accuracy on testing data $\textnormal{CVEfixes}_{test}^{year}$}
\label{tab:rq4_cvefixes_year}
\end{minipage}
\end{table*}

In RQ4, we explore different ways of creating test datasets, each of them capturing an important facet of transfer learning. \autoref{tab:rq4_big_vul_rand}, \autoref{tab:rq4_big_vul_year}, \autoref{tab:rq4_cvefixes_rand} and \autoref{tab:rq4_cvefixes_year} show the test sequence accuracies for all considered data splitting strategies. For each table, the first column lists the top-10 most common CWE IDs in each data split of the respective dataset. The second column shows the performance of the transfer learning model, which is a model trained on the large bug fix corpus, and then tuned with the vulnerability fix examples. The third column presents the performance of the model trained with the small dataset only, \ie only target domain training on each training split of different data splits. The last row of each table presents the test sequence accuracy on the whole test set on each data split.

% finding 1
From all tables, we can clearly see a large difference between the performance of transfer learning and target domain learning confirming the results of RQ2. The models that are only trained with the small vulnerability fix dataset (\ie target domain training), \minor{have a performance of 7.86\% on $\textnormal{Big-Vul}_{test}^{rand}$, 1.49\% on $\textnormal{Big-Vul}_{test}^{year}$, 11.29\% on $\textnormal{CVEfixes}_{test}^{rand}$ and 0.13\% on $\textnormal{CVEfixes}_{test}^{year}$}. They are all worse than the models trained with transfer learning whose performance are \minor{21.86\% on $\textnormal{Big-Vul}_{test}^{rand}$, 19.24\% on $\textnormal{Big-Vul}_{test}^{year}$, 22.73\% on $\textnormal{CVEfixes}_{test}^{rand}$ and 16.23\% on $\textnormal{CVEfixes}_{test}^{year}$}. For both data split strategies (random and time-based) the transfer learning model outperforms the target domain learning model showing that the result is independent of the data split. In other words, this is an additional piece of evidence about the superiority of transfer learning.

% finding 2: stable performance
Interestingly, the performance of models trained with target domain training only varies a lot between different data split strategies. It varies from \minor{0.13\%} in $\textnormal{CVEfixes}_{test}^{year}$ on the CVEfixes dataset to 11.29\% in $\textnormal{CVEfixes}_{test}^{rand}$. In other words, the performance of a vulnerability fixing system trained on a small dataset is unstable; it is highly dependent on how the data is divided into training, validation, and testing data. This has been observed in prior research as well \cite{li2007using}: the knowledge learned from a small dataset is often unreliable. On the other hand, transfer learning models have stable performance, staying in a high range from \minor{16.23\% on $\textnormal{CVEfixes}_{test}^{year}$ to 22.73\% on $\textnormal{CVEfixes}_{test}^{rand}$}.

When comparing the test sequence accuracy for each vulnerability type, \ie different CWE IDs, the transfer learning models also surpassed models trained only on the target domain. \minor{The only exception is CWE-190 in $\textnormal{CVEfixes}_{test}^{rand}$.} It may be that the nature of the fixes for the CWE varies over time such that transfer learning was more beneficial for the $\textnormal{Big-Vul}_{test}^{year}$ and $\textnormal{CVEfixes}_{test}^{year}$ splits. When comparing the overall performance on $\textnormal{Big-Vul}_{test}^{rand}$ and $\textnormal{CVEfixes}_{test}^{rand}$, target domain training versus transfer learning, the transfer learning model is still to be preferred.

We argue that splitting the vulnerability fix dataset based on time is the most appropriate split for evaluating a vulnerability repair system. By splitting the dataset based on time and having the newest vulnerabilities in the test set, we simulate a scenario where VRepair is trained on past vulnerability fixes and evaluated on future vulnerabilities. To this extent, we suggest that $\textnormal{Big-Vul}_{test}^{year}$ and $\textnormal{CVEfixes}_{test}^{year}$ are the most representative approximations of the performance of VRepair in practice.

\begin{mdframed}[nobreak=true]
Answer to RQ4: 
For the two considered data splitting strategies (random and time-based) transfer learning achieves more stable accuracies (\minor{16.23\% to 22.73\%} for transfer learning, and \minor{0.13\% to 11.29\%} for models only trained on the small vulnerability fix dataset). This validates the core intuition of VRepair: transfer learning overcomes the scarcity of vulnerability data for deep learning and yields reliable effectiveness.
\end{mdframed}

%%%%%%%%%%%%%%%%%%%%%%%%%%%%%%%%%%%%%%%%%%%%%%%%%%%%%%%%%%%%
%%%%%%%%%%%%%%%%%%%%%%%%%%%%%%%%%%%%%%%%%%%%%%%%%%%%%%%%%%%%
\section{Ablation Study}
\label{sec:ablation}

During the development of VRepair, we explored alternative architectures and data formats. To validate our explorations, we perform a systematic ablation study. \autoref{tab:ablation} highlights 8 comparisons that are of particular interest. The description column explains the way the model was varied which produced the results. All the models in the ablation study are evaluated on the Big-Vul dataset. ID 0 is our golden model of VRepair. We include it in the table for easier comparison against other architectures and dataset formats that we have tried. ID 1 summarizes the benefit of using beam search from the neural network model for our problem. In this comparison, the pass rate on our target dataset $\textnormal{Big-Vul}_{test}^{rand}$ increased from \minor{8.96\% with a single model output to 21.86\% with a beam size of 50}. ID 2 indicates the benefit of the Transformer architecture for our problem. Here we see that the Transformer model outperforms the bidirectional RNN model.

ID 3 highlights the importance of fault localization for model performance. When a model was trained and tested on raw vulnerable functions without any identification of the vulnerable line(s), we saw rather poor performance. When all vulnerable lines in the source file are identified, the model was much more likely to predict the correct patch to the function. Also, by localizing where the patch should be applied, there are fewer possible interpretations for the context matching to align with and this improves the viability of smaller context sizes. Given that we rely on fault localization for VRepair, labeling all possible vulnerable lines would be ideal and results in a \minor{23.9\%} test sequence accuracy. However, most static analysis tools will not provide this information. \minor{For example, Infer \cite{calcagno2011infer} only provides a single vulnerable line as output for each found vulnerability.}

% first line, single block, multi-line
If we limit our model to only predict changes for a contiguous block of lines (i.e, one or more lines after the erroneous line is identified), then the test sequence accuracy is \minor{28.39\%}. We note that single block repairs (which include single-line repairs) form 57.84\% of our dataset, so even with \minor{28.39\%} success, the single block model solves \minor{$28.39\% \times 57.84\% = 16.42\%$} of the test data, which is less than the \minor{21.86\%} our golden model solves. Ultimately, our model identifies only the first vulnerable line for input, but repairs may be done to lines after the identified line also. We consider it a reasonable assumption that a fault localization tool, \eg static analyzer or human, would tend to identify the first faulty line. In other words, we make no assumption if the first buggy vulnerable line is the only vulnerable line, meaning that VRepair still can fix multi line vulnerabilities, as we have seen in \autoref{lst:CVE-2016-9754}.

\begin{table}
\caption{A sample of ablation results over 8 hyperparameters.}\vspace{-.1cm} 
  \label{tab:ablation}
  \small
  \setlength\tabcolsep{2pt}
  \centering
  \begin{tabular}{@{}lllll@{}}
    \toprule
 ID & & Description & & Results \\
    \midrule
0 & & \textbf{VRepair} & & \textbf{21.86\%} \\ \hline
1 & & Beam width 1 & & 8.96\% \\
  & & Beam width 10 & & 17.61\% \\
  & & Beam width 50 (\textbf{VRepair}) & & \textbf{21.86\%} \\ \hline
2 & & RNN seq2seq & & 17.14\% \\
  & & Transformer seq2seq (\textbf{VRepair}) & & \textbf{21.86\%} \\ \hline
3 & & No vulnerable line identifier  & & 10.99\% \\
  & & All vulnerable lines ID'd & & 23.90\% \\
  & & Single block ID'd & & 28.39\% \\
  & & First vulnerable line ID'd (\textbf{VRepair}) & & \textbf{21.86\%} \\ \hline
4 & & Learn rate 0.0001 (\textbf{VRepair}) & & \textbf{21.86\%} \\
  & & Learn rate 0.0005 & & 0.00\% \\
  & & Learn rate 0.00005 & & 19.97\% \\ \hline
5 & & Hidden size 1024 (\textbf{VRepair}) & & \textbf{21.86\%} \\
  & & Hidden size 512 & & 20.28\% \\ \hline
6 & & Layers 6 (\textbf{VRepair}) & & \textbf{21.86\%} \\ 
  & & Layers 4 & & 18.87\% \\ \hline
7 & & Dropout 0.1 (\textbf{VRepair}) & & \textbf{21.86\%} \\
  & & Dropout 0.0 & & 17.45\% \\  \hline
8 & & Vocabulary size 5000 (\textbf{VRepair}) & & \textbf{21.86\%} \\ 
  & & Vocabulary size 2000 & & 17.61\% \\
  & & Vocabulary size 10000 & & 21.86\% \\
     \bottomrule
  \end{tabular}
\end{table}

IDs 4 to 7 show our ablation of the key hyperparameters in our golden model. ID 8 shows why our vocabulary size is set to a rather low value of 5000, which is done thanks to using the copy mechanism. We clearly see that by using the copy mechanism, the model can handle the out-of-vocabulary problem well with a low vocabulary size \minor{of 5000}.

\section{Related Work}

% Related work about vulnerability fixing using ML
\subsection{Vulnerability Fixing with Learning}

% Catch all for vulnerability detection and automatic program repair.
% There is a significant amount of research on vulnerability detection \cite{cao2021bgnn4vd, li2018vuldeepecker, russell2018automated, harer2018automated}. For vulnerability detection based on machine learning, we refer the reader to surveys on approaches using deep neural network to detect software vulnerabilities \cite{lin2020software} and using artificial intelligence techniques to prevent vulnerabilities \cite{kommrusch2019AIVuln}. In the rest of this subsection, we discuss previous works that go beyond detection, and also do some kind of automatic vulnerability repair. For a generic overview on automatic program repair, we refer the reader to  \cite{monperrus:hal-01956501}.

We include related work that fixes vulnerabilities with some kind of learning, meaning that the system should learn fix patterns from a dataset, instead of generating repairs from a set of pre-defined repair templates.

% Vurle: Automatic vulnerability detection and repair by learning from examples
Vurle is a template based approach to repair vulnerability by learning from previous examples \cite{ma2017vurle}. They first extract the edit from the AST diff between the buggy and fixed source code. All edits are then categorized into different edit groups to generate repair templates. To generate a patch, Vurle identifies the best matched edit group and applies the repair template. Vurle is an early work that does not use any deep learning techniques and is only trained and evaluated on \num{279} vulnerabilities. In contrast, VRepair is trained and evaluated on \num{3754} vulnerabilities and is based on deep learning techniques.

% Learning to repair software vulnerabilities with generative adversarial networks
Harer \etal \cite{harer2018learning} proposed using generative adversarial networks (GAN), to repair software vulnerabilities. They employed a traditional neural machine translation (NMT) model as the generator to generate the examples to confuse the discriminator. The discriminator task is to distinguish the NMT generated fixes from the real fixes. The trained GAN model is evaluated on the SATE IV dataset \cite{okun2013report} consisting of synthetic vulnerabilities. Although Harer \etal trained and evaluated their work on a dataset of \num{117 738} functions, the main limitation is that the dataset is fully synthetic. In contrast, our results have better external validity, because they are only based on real world code data.

% SeqTrans: Automatic Vulnerability Fix via Sequence to Sequence Learning
SeqTrans is the closest related work \cite{chi2020seqtrans}. SeqTrans is a Transformer based seq2seq neural network with attention and copy mechanisms that aims to fix Java vulnerabilities. Similar to VRepair, they also first train on a bug fix corpus, and then fine-tune on a vulnerability fix dataset. Their input representation is the vulnerable statement, and the statements that defined the variables used in the vulnerable statement. To reduce the vocabulary, the variable names in the buggy and fixed methods are renamed and they use BPE to further tokenize the tokens. VRepair is different from SeqTrans in that we target fixing C vulnerabilities instead of Java vulnerabilities. Importantly, we utilize the copy mechanism to deal with tokens outside the training vocabulary and not BPE. Our evaluation is done based on two independent vulnerability fix datasets to increase the validity. VRepair's code representation is also different, and allows us to represent multi-line fixes in a compact way. We have shown in \autoref{lst:CVE-2016-9754} that we are able to fix multi-line vulnerabilities, while SeqTrans focused on single statement vulnerabilities.

\subsection{Vulnerability Fixing without Learning}

We include related work that fixes vulnerabilities without learning. Usually, these works do so by having a pool of pre-defined repair templates or using different program analysis techniques to detect and repair the vulnerability.

% Using safety properties to generate vulnerability patches
Senx is an automatic program repair method that generates vulnerability patches using safety properties \cite{huang2019using}. A safety property is an expression that can be mapped to variables in the program, and it corresponds to a vulnerability type. It uses the safety property to identify and localize the vulnerability and then Senx generates the patch. In the implementation, three safety properties are implemented: buffer overflow, bad cast, and integer overflow. For buffer overflow and bad cast, Senx generates a patch where the error handling code is called. But for integer overflow, Senx will generate a patch where the vulnerability is actually fixed.

% The Mayhem Cyber Reasoning System
Mayhem is a cyber reasoning system that won the DARPA Cyber Grand Challenge in 2016 \cite{avgerinos2018mayhem}. It is able to generate test cases that expose a vulnerability and to generate the corresponding binary patch. The patches are based on runtime property checking, \ie assertions that are likely to be true for a correctly behaving program and false for a vulnerable program. To avoid inserting many unnecessary checks, Mayhem uses formal methods to decide which runtime checks to add. 

% Fuzzbomb: Autonomous cyber vulnerability detection and repair
Fuzzbuster is also a cyber reasoning system that has participated in the DARPA Cyber Grand Challenge \cite{musliner2015fuzzbomb}. It can find security flaws using symbolic execution and fuzz testing, along with generating binary patches to prevent vulnerability. The patches typically shield the function from malicious input, such as a simple filter rule that blocks certain inputs. 

% Beyond Tests: Program Vulnerability Repair via Crash Constraint Extraction
ExtractFix is an automated program repair tool that can fix vulnerabilities that can cause a crash \cite{gao2020beyond}. It works by first capturing constraints on inputs that must be satisfied, the constraints capturing the properties for all possible inputs. Then, they find the candidate fix locations by using the crash location as a starting point, and use control/data dependency analysis to find candidate fix locations. The constraints, together with the candidate fix locations, are used to generate a patch such that the constraints cannot be violated at the crash location in the patched program.

% Memfix: static analysis-based repair of memory deallocation errors for c
MemFix is a static analysis based repair tool for fixing memory deallocation errors in C programs \cite{lee2018memfix}, \eg memory leak, double free, and use-after-free errors. It does so by first generating candidate patches for each allocated object using a static analyzer. The correct patches are the candidate patches that correctly deallocate all object states. It works by reducing the problem into an exact cover problem and using an SAT solver to find the solution.

% Safe Memory-Leak Fixing for C Programs
LeakFix is an automatic program repair tool for fixing memory leaks in C programs \cite{gao2015safe}. It first builds the control flow graph of the program and uses intra-procedural analysis to detect and fix memory leaks. The generated fix is checked against a set of procedures that ensures the patch does not interrupt the normal execution of the program.

% Automatically patching errors in deployed software
ClearView is an early work that can protect deployed software by automatically patching vulnerabilities \cite{perkins2009automatically}. To do so, ClearView first observes the behavior of software during normal execution and learns invariants that are always satisfied. It uses the learned invariant to detect and repair failures. The patch can change register values and memory locations, or change the control flow. It has been able to resist 10 attacks from an external Red Team.

% Intrepair: Informed repairing of integer overflows
IntRepair is a repair tool that can detect, repair, and validate patches of integer overflows \cite{muntean2019intrepair}. It uses symbolic execution to detect integer overflows by comparing the execution graphs with three preconditions. Once an integer overflow is detected, a repair pattern is selected and applied. The resulting patched program is executed again with symbolic execution to check if the integer overflow is repaired.

% Diagnosis and emergency patch generation for integer overflow exploits
SoupInt is a system for diagnosing and patching integer overflow exploits \cite{wang2014diagnosis}. Given an attack instance, SoupInt will first determine if the attack instance is an integer overflow exploit using dynamic data flow analysis. Then, a patch can be generated with different policies. It can either change the control flow or perform a controlled exit.

VRepair is different than all these vulnerability fix systems since it is a learning based system. The major difference is that VRepair is not targeting a specific vulnerability, rather it is able to fix multiple types of vulnerabilities, as seen in \autoref{tab:rq2_big_vul} and \autoref{tab:rq2_cvefixes}. VRepair is also designed to be able to learn arbitrary vulnerability fixes, rather than having to manually design a repair strategy for each type of vulnerability.

\subsection{Vulnerability Datasets}

% A C/C++ Code Vulnerability Dataset with Code Changes and CVE Summaries
Big-Vul is a C/C++ code vulnerability dataset collected from open sourced GitHub projects \cite{fan2020ac}. It contains \num{3754} vulnerabilities with 91 different vulnerability types extracted from 348 GitHub projects. Each vulnerability has a list of attributes, including the CVE ID, CWE ID, commit ID, \etc This dataset can be used for vulnerability detection, vulnerability fixing, and analyzing vulnerabilities.

% CVEfixes: Automated Collection of Vulnerabilities and Their Fixes from Open-Source Software
CVEfixes \cite{bhandari2021cvefixes} is a vulnerability fix dataset based on CVE records from National Vulnerability Database (NVD) . The vulnerability fixes are automatically gathered from the associated open-source repositories. It contains CVEs up to 9 June 2021, with 5365 CVE records in 1754 projects.

% A ground-truth dataset of real security patches
Reis \& Abreu collected a dataset of security patches by mining the entire CVE details database \cite{Reis2021AGD}. This dataset contains 5942 security patches gathered from 1339 projects with 146 different vulnerability types in 20 languages. They also collected 110k non-security related commits which are useful in training a system for identifying security relevant commits.

% Vuldeepecker: A deep learning-based system for vulnerability detection
A vulnerability detection system, VulDeePecker \cite{li2018vuldeepecker}, created the Code Gadget Database. The dataset contains \num{61638} C/C++ code gadgets, in which \num{17725} of them are labeled as vulnerable and the remaining \num{43913} code gadgets are labeled as not vulnerable. The Code Gadget Database only focuses on two types of vulnerability categories: buffer error (CWE-119) and resource management error (CWE-399). By contrast, the Big-Vul that is used in this paper contains vulnerabilities with 91 different CWE IDs.

% SeqTrans: Automatic Vulnerability Fix via Sequence to Sequence Learning
Ponta \etal created a manually curated dataset of Java vulnerability fixes \cite{ponta2019manually} which has been used to train SeqTrans, a vulnerability fixing system \cite{chi2020seqtrans} presented above. The dataset has been collected through a vulnerability assessment tool called "project KB", which is open sourced. In total, Ponta's dataset contains 624 vulnerabilities collected from 205 open sourced Java projects.

% Learning to repair software vulnerabilities with generative adversarial networks
SATE IV is a vulnerability fix dataset originally used to evaluate static analysis tools on the task of finding security relevant defects \cite{okun2013report}. SATE IV consists of \num{117738} synthetic C/C++ functions with vulnerabilities spanning across 116 CWE IDs, where \num{41171} functions contain a vulnerability and \num{76567} do not.

% VulinOSS: a dataset of security vulnerabilities in open-source systems
VulinOSS is a vulnerability dataset gathered from open source projects \cite{gkortzis2018vulinoss}. The dataset is created from the vulnerability reports of the National Vulnerability Database. They manually assessed the dataset to remove any projects that do not have a publicly available repository. In total, the dataset contains \num{17738} vulnerabilities from \num{153} projects across \num{219} programming languages. 

Cao \etal collected a C/C++ vulnerability dataset from GitHub and the National Vulnerability Database, consisting of \num{2149} vulnerabilities \cite{cao2021bgnn4vd}. They used the dataset to train a bi-directional graph neural network for a vulnerability detection system.

% final summary of this subsection
% In our work, we choose to use Big-Vul and CVEfixes as our vulnerability fix datasets because they are the most recent dataset with C/C++ vulnerabilities, and it has the largest number of C/C++ vulnerabilities.

\subsection{Machine Learning on Code}

Here we present related works that use machine learning on source code. In general, we refer the reader to the survey by Allamanis \etal for a comprehensive overview on the field \cite{allamanis2018survey}.

% Deepfix: Fixing common c language errors by deep learning
One of the first papers that used seq2seq learning on source code is DeepFix \cite{gupta2017deepfix}, which is about fixing compiler errors. They encode the erroneous program by replacing variable names with a pre-defined pool of identifiers to reduce the vocabulary. The program is then tokenized, and a line number token is added for each line, so that the output can predict a fix for a single code line. C programs written by students are encoded and used to train DeepFix, and it was able to fix 27\% of all compiler errors completely and 19\% of them partially.

% An empirical study on learning bug-fixing patches in the wild via neural machine translation
Tufano \etal investigated the possibility of using seq2seq learning to learn bug fixing patches in the wild \cite{tufano2019empirical}. Similar to our approach, they collected a bug fix corpus by filtering commits from GitHub based on the commit message. The input is the buggy function code, where identifiers are replaced with a general name, such as STRING\_X, INT\_X, and FLOAT\_X. Then, they trained a seq2seq model where the output is the fixed function code. They found that seq2seq learning is a viable approach to learn how to fix code, but found that the model's performance decreased with the output length.

% Sequencer: Sequence-to-sequence learning for end-to-end program repair
SequenceR learned to fix single line Java bugs using seq2seq learning \cite{chen2019sequencer}. The input to SequenceR is the buggy class, where unnecessary context is removed, such as the function body of non-buggy functions. The input also encodes the fault localization information by surrounding the suspicious line with <START\_BUG> and <END\_BUG>. The output is the code line to replace the suspicious line. They used the copy mechanism to deal with the vocabulary problem, instead of renaming identifiers. SequenceR was evaluated on Defects4J \cite{just2014defects4j} and was able to fix 14 out of 75 bugs. \minor{The biggest difference compared to this previous work is the usage of   
transfer learning in VRepair, whereas SequenceR is only trained on a single domain, which is the bug fix task. Additionally, VRepair is able to generate multi-line patches as opposed to the single line fixes of SequenceR.}

% CoCoNuT: Combining context-aware neural translation models using ensemble for program repair
CoCoNut is an approach that combines multiple seq2seq models to repair bugs in Java programs \cite{lutellier2020coconut}. They used two encoders to encode the buggy program; one encoder generates a representation for the buggy line, and another encoder generates a representation for the context. These two representations are then merged to predict the bug fix. They trained multiple different models and used ensemble learning to combine the predictions from all models.

% DLfix: Context-based code transformation learning for automated program repair
DLFix is an automated program repair tool for fixing Java bugs \cite{li2020dlfix}. It is different from other approaches in that it uses tree-based RNN and has two layers. The first layer is a tree-based RNN that encodes the AST that surrounds the buggy source code, which is passed to the second layer. The second layer takes the context vector and learns how to transform the buggy sub-tree. It will generate multiple patches for a single bug, and it deploys a character level CNN to rank all the generated patches.

% code2seq: Generating Sequences from Structured Representations of Code
Code2Seq \cite{alon2018code2seq} uses AST paths to represent a code snippet for the task of code summarization. An AST path is a path between two leaf nodes in the AST. They sample multiple AST paths and combine them using a neural network to generate sequences such as function name, code caption, and code documentation. They found that by using AST paths, Code2Seq can achieve better performance than seq2seq neural network and other types of neural networks.

% CODIT: Code Editing with Tree-Based Neural Models
CODIT is a tree-based neural network to predict code changes \cite{chakraborty2020codit}. The neural network takes the AST as input, and generates the prediction in two steps. The first step is a tree-to-tree model that predicts the structural changes on the AST, and the second step is the generation of code fragments. They evaluate CODIT on real bug fixes and found that it outperforms seq2seq alternatives.

% Learning to represent edits
Yin \etal worked on the problem of learning distributed representation of edits \cite{yin2018learning}. The edits they learned are edits on Wikipedia articles and edits on GitHub C\# projects. They found that similar edits are grouped together when visualizing the edit representation, and that transferring the neural representation of edits to to a new context is indeed feasible.

The major difference between VRepair and these works is that VRepair is targeting vulnerabilities. In this work, we evaluate VRepair with the most notable vulnerabilities which have a CVE ID; they are all confirmed vulnerabilities reported by security researchers. These vulnerabilities are essentially different from related works which consider functional bugs, for example from Defects4J \cite{just2014defects4j}.

% Related works about transfer learning in SE
\subsection{Transfer Learning in Software Engineering}

% Exploring the Possibilities of Applying Transfer Learning Methods for Natural Language Processing in Software Development
To the best of our knowledge, there are only a few works that use transfer learning in the software engineering domain, and none of them use it for generating code fixes. Recently, Ding has done a comprehensive study on applying transfer learning to different software engineering problems \cite{Wei2021Master}, such as code documentation generation and source code summarization. He found that transfer learning improves performance on all problems, especially when the dataset is tiny and could be easily overfitted. In our work, we deploy transfer learning for vulnerability fixing and show that it also improves accuracy. Also, we show that our model trained with transfer learning has a more stable and superior performance compared to training on the small dataset.

% Application of Seq2Seq Models on Code Correction
Huang, Zhou, and Chin used transfer learning to avoid the problem of having a small dataset for the error type classification task, \ie predict the vulnerability type \cite{huang2020application}. They trained a Transformer model on the small dataset and achieved 7.1\% accuracy. When training first on a bigger source dataset and tuned afterward on the same dataset, Huang \etal \cite{huang2020application} managed to get 69.1\% accuracy. However, their work is not about transfer learning and therefore the transfer learning experiment was relatively simple and short. In our work, we conduct multiple experiments to show the advantages of using transfer learning for vulnerability fixing. 

% Code smell detection by deep direct-learning and transfer-learning
Sharma \etal have applied transfer learning on the task of detecting code smell \cite{sharma2021code}. They train the models on C\# code and use them to detect code smells with Java code examples and vice versa. They found that such models achieved similar performance to models directly trained on the same program language. This is different from our work; we observed that transfer learning improved the performance and made the performance more stable.

% Easy-to-Deploy API Extraction by Multi-Level Feature Embedding and Transfer Learning
Ma \etal used character, word and sentence-level features from the input text to recognize API uses. They adopted transfer learning to adapt a neural model trained on one API library to another API library. They found that the more similar the API libraries are, the more effective transfer learning is. Overall, this related literature, together with our paper, show the applicability and benefits of using transfer learning in software engineering.

\section{Conclusion}

In this paper, we have proposed VRepair, a novel system for automatically fixing C vulnerabilities using neural networks. To tackle the problem of having only small vulnerability fix datasets, our key insight is to use transfer learning: we first train VRepair with a big bug-fix corpus, and then we fine-tune on a curated vulnerability fix dataset. 

We have performed a series of original and large scale experiments. In our experiments, we found that VRepair's transfer learning outperforms a neural network that is only trained on the small vulnerability fix dataset, attaining \minor{21.86\% accuracy instead of 7.86\%} on Big-Vul dataset, and \minor{22.73\% accuracy instead of 11.29\%} on CVEfixes dataset. This result shows that transfer learning is a promising way to address the small dataset problem in the domain of machine learning for vulnerability fixing. Put in another way, our experiments show the knowledge learned from the bug fixing task can be transferred to the vulnerability fixing task, which has never been studied before to the best of our knowledge.

In the future, we would like to explore the possibility of using an even larger source domain dataset. We believe that one can achieve results similar to those of GPT-3 -- a massive, generic natural language model with 175 billion parameters \cite{brown2020language}. In the context of software engineering, we envision that we could train a single model on all code changes from GitHub, not just bug fixes, and tune it on tasks such as code comment generation, function name prediction, and vulnerability fixing. \minor{It is also possible to enlarge the considered target domain datasets, for example with vulnerabilities reported by issue tracking tools and various static analyzers.}

\section*{Acknowledgments}
This project is partially financially supported by the Swedish Foundation for Strategic Research (SSF). This work was supported in part by the U.S. National Science Foundation award CCF-1750399. The computations was enabled by resources provided by the Swedish National Infrastructure for Computing (SNIC), partially funded by the Swedish Research Council through grant agreement no. 2018-05973.

\ifCLASSOPTIONcaptionsoff
  \newpage
\fi



%\bibliographystyle{IEEEtran}

\printbibliography


%\begin{IEEEbiography}{Zimin Chen}
%\end{IEEEbiography}

%\begin{IEEEbiographynophoto}{Steve Kommrusch}

%\newpage

%\begin{IEEEbiographynophoto}{Martin Monperrus}


%\vfill


\end{document}